\documentclass[lettersize,journal]{IEEEtran}
\usepackage{amsmath,amsfonts}
\usepackage{algorithmic}
\usepackage{algorithm}
\usepackage{array}
\usepackage{textcomp}
\usepackage{stfloats}
\usepackage{url}
\usepackage{cite}

\usepackage{amssymb}
\usepackage{float}
\usepackage{graphicx}
\usepackage{subcaption}
\usepackage{amsmath}
\usepackage{enumerate}
\usepackage{booktabs} 
\usepackage{setspace}
\usepackage{tabularx}
\usepackage{enumerate}
\begin{document}

\title{PPBFL: A Privacy Protected Blockchain-based Federated Learning Model}
%
\author{Yang~Li,
	Chunhe~Xia,
	Wanshuang~Lin, 
	and~Tianbo~Wang,~\IEEEmembership{Member,~IEEE,}
	\thanks{Yang Li is with the Key Laboratory of Beijing Network Technology, Beihang University, Beijing 100191, China (e-mail: johnli@buaa.edu.cn).}
	\thanks{Chunhe Xia is with the Key Laboratory of Beijing Network Technology, Beihang University, Beijing 100191, China, and also with the Guangxi 		Collaborative Innovation Center of Multi-Source Information Integration and Intelligent Processing, Guangxi Normal University, Guilin 541004, China 		(e-mail: xch@buaa.edu.cn).}
	\thanks{Wanshuang Lin is with the School of Cyber Science and Technology, Beihang University, Beijing 100191, China (e-mail: linws@buaa.edu.cn).}%
	\thanks{Tianbo Wang is with the School of Cyber Science and Technology, Beihang University, Beijing 100191, China, and also with the Shanghai Key Laboratory 	of Computer Software Evaluating and Testing, Shanghai 201112, China (e-mail: wangtb@buaa.edu.cn). } 
	\thanks{(\textit{Corresponding author:}Tianbo Wang.)}
	\thanks{Manuscript received April 19, 2005; revised August 26, 2015.}}

\markboth{Journal of \LaTeX\ Class Files,~Vol.~14, No.~8, August~2021}%
{Shell \MakeLowercase{\textit{et al.}}: A Sample Article Using IEEEtran.cls for IEEE Journals}


\maketitle
\begin{abstract}

With the rapid development of machine learning and a growing concern for data privacy, federated learning has become a focal point of attention. However, attacks on model parameters and a lack of incentive mechanisms hinder the effectiveness of federated learning. Therefore, we propose A Privacy Protected Blockchain-based Federated Learning Model (PPBFL) to enhance the security of federated learning and encourage active participation of nodes in model training. Blockchain technology ensures the integrity of model parameters stored in the InterPlanetary File System (IPFS), providing protection against tampering. Within the blockchain, we introduce a Proof of Training Work (PoTW) consensus algorithm tailored for federated learning, aiming to incentive training nodes. This algorithm rewards nodes with greater computational power, promoting increased participation and effort in the federated learning process. A novel adaptive differential privacy algorithm is simultaneously applied to local and global models. This safeguards the privacy of local data at training clients, preventing malicious nodes from launching inference attacks. Additionally, it enhances the security of the global model, preventing potential security degradation resulting from the combination of numerous local models. The possibility of security degradation is derived from the composition theorem. By introducing reverse noise in the global model, a zero-bias estimate of differential privacy noise between local and global models is achieved. Furthermore, we propose a new mix transactions mechanism utilizing ring signature technology to better protect the identity privacy of local training clients. Security analysis and experimental results demonstrate that PPBFL, compared to baseline methods, not only exhibits superior model performance but also achieves higher security.

\end{abstract}


\section{Introduction}

%

Machine learning (ML) applications have become increasingly widespread in people's work and daily lives \cite{Carvalho2019MachineLI}, profoundly transforming both work and lifestyle. The application of various ML models provides more intelligent assistance, offering new ways for image, speech, and text recognition and processing. The progress in ML over the years has greatly enhanced work efficiency and the quality of life. However, ML is a double-edged sword. From a positive perspective, the powerful learning ability of ML for data has made it crucial in areas such as vision detection \cite{Ouhami2021ComputerVI}, text generation \cite{Naithani2022RealizationON}, and intelligent transportation \cite{Zhang2011DataDrivenIT}. On the negative side, as ML model training requires extensive real datasets, incidents of privacy data leaks have become more frequent. Thus, safeguarding the privacy of ML training data has become a new focus\cite{Rigaki2020ASO}.

Federated learning (FL) is an emerging framework for training ML models while protecting the privacy of device data\cite{Yang2019FederatedML}, with broad application prospects. In the context of the increasing prevalence of artificial intelligence, FL has emerged to meet the growing demand for higher data privacy. In particular, FL shares only model parameters during the model training process, while training data remains stored locally on client devices. This means that we can better protect the privacy of training data while achieving model training.

%

However, there are two inevitable challenges in FL: \textit{inference-attacks} \cite{Zhang2021ASO} and  \textit{client-incentives} \cite{Tu2021IncentiveMF}. \textit{Inference-attacks}  may arise from malicious clients or the aggregation server. Although we only share model parameters in FL, these parameters contain certain training data information. Analyzing the model parameters shared by benign clients may lead to the inference of the data distribution or the presence of specific data in their training datasets.  \textit{Client-incentives} refer to the scenario where, during the FL process, local training clients need to collaborate to accomplish the FL task. However, instances of "lazy" clients may occur, where clients may not utilize all computational resources to complete the FL task or may fail to train for the specified number of rounds, resulting in poor performance of the local model. Client incentives aim to motivate local training clients to actively participate in FL tasks.

Benefiting from FL's outstanding performance in data privacy and security, many researchers have started incorporating homomorphic encryption \cite{Fang2021PrivacyPM} and differential privacy \cite{Ouadrhiri2022DifferentialPF} to protect data. Researchers initially employed homomorphic encryption to encrypt transmitted model parameters. After performing homomorphic operations at the aggregation server, the computed results were returned to the local training clients. The local clients then decrypted the results to obtain the global model. In contrast, differential privacy methods involve centralizing data on the server, which adds differential privacy noise and provides services to data queryers. Unfortunately, these two methods fall short of addressing the issue of \textit{inference-attacks} . Firstly, homomorphic encryption requires significant computational resources, which may not meet the varied computing capabilities of local training clients in FL. Secondly, the differential privacy approach necessitates local training clients sending plaintext to the server, without guaranteeing the complete trustworthiness of the server. Lastly, traditional differential privacy methods, due to the randomness and uncontrollability of the added noise in terms of size and position, may even impact model performance.
In summary, it is crucial to design new methods to safeguard the privacy of FL training nodes and training data. This promotes the enthusiasm of local training nodes to participate in model training, enhancing the overall performance of the FL global model.

%
%
%
To overcome the aforementioned obstacles, we propose a Privacy Protected Blockchain-based Federated Learning Model (PPBFL). This model adds differential privacy noise to both local training clients and global model aggregation clients. The former protects the privacy of the local model's training data, while the latter prevents the decrease in differential privacy security, as indicated by the composition theorem, when too many local models with local differential privacy noise are added in FL. Security analysis and experimental results demonstrate that our proposed model enhances data security while ensuring model performance.

Our contributions can be summarized as follows:
\begin{itemize}
	\item We introduce the PPBFL model to address the issue of inference attacks. We design a dual local differential privacy mechanism to protect the privacy of benign clients' data from both the server side and malicious client side. We propose a novel adaptive local differential privacy noise addition method, reducing the required differential privacy noise while satisfying data privacy security.
	\item We propose a mixing CID mechanism based on ring signatures to protect the identity privacy of local training nodes.
	\item We introduce a consensus algorithm based on the federated training work of local training clients, incentive's them to participate in FL training.
	\item We formally prove that our differential privacy scheme satisfies $\epsilon$-LDP security and introduces zero bias when estimating average weights. Experiments show that our proposed differential privacy scheme achieves better model performance while preserving $\epsilon$-LDP security.
\end{itemize}

In conclusion, our PPBFL model addresses \textit{inference-attacks}  through innovative differential privacy methods, and identity privacy protection mechanisms, incentive clients with consensus algorithms. The formal security proof and experimental results validate the effectiveness of our proposed approach in enhancing both model performance and data security in FL.


The remainder of this article is structured as follows. In Section \ref{RELATED WORK}, we provide a summary and review of relevant literature. Section \ref{Preliminaries} introduces the background knowledge. Subsequently, in Section \ref{Privacy Protected Blockchain-based Federated Learning model}, we delve into the discussion of model design and architecture. Section \ref{ Security Analysis} presents a security analysis of the proposed model. The performance of the model is evaluated in Section \ref{Experimental Results}. Lastly, Section \ref{Conclusion} serves as the conclusion for this paper.

\section{RELATED WORK}
\label{RELATED WORK}
%

This section provides a comprehensive review of relevant literature. We start by briefly introducing the work on model parameter privacy protection in FL. Subsequently, we delve into the state-of-the-art approaches that employ differential privacy for safeguarding FL and the efforts in incorporating incentive mechanisms into FL.

Commonly used methods for preserving privacy of model parameters in FL include homomorphic encryption, secure multi-party computation, and differential privacy. In homomorphic encryption, local training clients encrypt their model parameters using homomorphic encryption and send the encrypted parameters to the aggregation server. The server performs homomorphic calculations on the received encrypted  parameters to obtain the aggregated encrypted global model parameters. The server then sends the global model parameters back to the local training clients, who decrypt the global model parameters locally and proceed with the next round of training. Secure multi-party computation usually employs homomorphic encryption techniques, allowing multiple FL participants to jointly compute the gradient updates of the model parameters without sharing the actual parameter values. Differential privacy, on the other hand, introduces a small amount of noise to the model parameters, providing a privacy guarantee without compromising model performance. It ensures privacy of the model parameters within a certain privacy budget, allowing the model to perform securely.
%
%

Numerous studies have been conducted on FL and differential privacy technologies.
Kang Wei et al. introduced a novel framework based on the concept of differential privacy (DP), termed Noise Before Aggregation FL (NbAFL), where artificial noise is added to the client's parameters before aggregation \cite{Wei2019FederatedLW}.
Stacey Truex et al. proposed LDP-Fed, which provides privacy guarantees in the form of local differential privacy (LDP) \cite{Truex2020LDPFedFL}. In LDP-Fed, two novel methods are designed and developed. The LDP module of LDP-Fed offers formal differential privacy guarantees for repeatedly collecting model training parameters on private datasets from multiple individual participants during joint training of large-scale neural networks. 
%
Yang Zhao et al. propose the integration of FL and Local Differential Privacy (LDP) to facilitate the realization of ML models in crowdsourced applications \cite{Zhao2020LocalDP}. Specifically, four LDP mechanisms are introduced to perturb gradients generated by vehicles. The three proposed output mechanisms incorporate three different output possibilities to achieve high accuracy under limited privacy budgets. The likelihood of the three outputs can be encoded using two bits, effectively reducing communication costs. 
Antonious M. Girgis et al. address a distributed empirical risk minimization (ERM) optimization problem with considerations for communication efficiency and privacy requirements \cite{Girgis2021ShuffledMO}, as motivated by the federated learning (FL) framework . A distributed communication-efficient and locally differentially private stochastic gradient descent (CLDP-SGD) algorithm is proposed, and its trade-offs between communication, privacy, and convergence are analyzed. 
%
Bin Jia et al. have designed an application model for blockchain-based federated learning in the Industrial Internet of Things (IIoT) and have formulated a data protection aggregation scheme based on the proposed model \cite{Jia2021BlockchainEnabledFL}. 
Xicong Shen et al. have developed a performance-enhanced DP-based Federated Learning (PEDPFL) algorithm\cite{Shen2022PerformanceEnhancedFL}. The paper introduces a classifier perturbation regularization method to enhance the robustness of the trained model to differential privacy-induced noise. 
%
%
Laraib Javed et al. describe a secure and reliable data sharing architecture and semantic approach based on blockchain, Local Differential Privacy (LDP), and Federated Learning (FL) \cite{Javed2022ShareChainBM}. The proposed framework establishes a trustless environment where data owners no longer need to trust a central controller.
Yuntao Wang et al. design a novel block structure, new transaction types, and credit-based incentives in PF-PoFL \cite{Wang2022APP}. PF-PoFL allows efficient outsourcing of artificial intelligence (AI) tasks, collaborative mining, model evaluation, and reward allocation in a fully decentralized manner while resisting deception and Sybil attacks. 
%
%
%
%
%

Similarly, there is considerable research on how to incentivize local training clients to perform better in federated learning.
Yufeng Zhan et al. investigated incentive mechanisms in federated learning to motivate edge nodes to participate in model training \cite{Zhan2020ALI}. Specifically, they designed an incentive mechanism based on Deep Reinforcement Learning (DRL) to determine the optimal pricing strategy for the parameter server and the optimal training strategy for edge nodes.
Han Yu et al. proposed a Federated Learning Incentive (FLI) profit-sharing scheme \cite{Yu2020AFI}. The scheme aims to jointly maximize collective utility while minimizing inequality among data owners, dynamically allocating a given budget among data owners in a context-aware manner in the federated setting.
Yufeng Zhan et al. examined incentive mechanism design in federated learning \cite{Zhan2021ASO}, introducing a classification of existing incentive mechanisms in federated learning. They further conducted an in-depth discussion by comparing and contrasting different methods. 
Wen Sun et al. considered dynamic digital twins and federated learning in a space-ground network, where drones serve as aggregators and ground clients capture the dynamically evolving network through digital twin-based collaborative training models \cite{Sun2022DynamicDT}.
%
%
%
Han Yu et al. proposed FL Incentivizer (FLI) \cite{Yu2020ASI}. It dynamically allocates a given budget among data owners in federated settings in a context-aware manner. 
Yongheng Deng et al. introduced a new framework, FAIR \cite{Deng2022ImprovingFL}, for Federated Learning with Quality Assurance. FAIR integrates three main components: 1) Learning Quality Estimation; 2) Quality-Aware Incentive Mechanism; and 3) Automatic Weighted Model Aggregation.
Yanru Chen et al. integrated reputation-based and payment-based incentive measures \cite{Chen2022DIMDSDI}, introducing "reputation coins" as cryptocurrency for data-sharing transactions to encourage users to participate honestly in the data-sharing process based on federated learning. 
Tianle Mai et al. designed a dual auction mechanism for the FL service market \cite{Mai2022AutomaticDM}, where trained models can automatically trade between AIoT devices and FL platforms. 


However, more effective methods for preserving model parameter privacy and encouraging the engagement of training clients in federated learning are still areas that require further research. Moreover, there is a need for more attention to identity protection for local training clients in federated learning.

\section{Preliminaries}
\label{Preliminaries}
\subsection{ Blockchain}



Blockchain technology was initially proposed by Satoshi Nakamoto in 2008 within Bitcoin \cite{Nakamoto2008BitcoinAP}. It sequentially connects blocks storing transaction records in a chain-like fashion, utilizing cryptographic algorithms to ensure the immutability and authenticity of the blocks \cite{Zheng2018BlockchainCA, Berdik2021ASO, Zhou2020SolutionsTS}. A block is divided into two parts: the block header and the block body. The block header contains identifiers for the previous, current, and next blocks, a timestamp, and the Merkle root of the transactions forming a Merkle tree within the block body. The block body comprises transactions generated in the blockchain network during a specific time period. The transactions are pairwise hashed, ultimately producing a hash that serves as the Merkle root. This root is stored in the block header, and if transactions within the block are tampered with, the Merkle root will change. Consistency in the Merkle root ensures the tamper resistance of transactions within the block. A hash generated by comprehensively calculating the block content and timestamp serves as the block identifier. Blocks are connected through identifiers, and if the block content changes, the identifier changes accordingly, using the unidirectionality of hash functions to guarantee identifier uniqueness and block tamper resistance. In addition to tamper resistance, blockchain also possesses features such as public transparency, traceability of transaction records, and collaborative maintenance. Transaction records on the blockchain are broadcasted throughout the network, allowing nodes in the network to inspect the content of transactions within each block. Transaction records are stored in the blockchain, ensuring traceability.

Consensus algorithms, as a critical component of blockchain, ensure data consistency and consensus security \cite{Zheng2017AnOO, Xiao2019ASO}. Nodes in the blockchain network compete for mining rights based on rules set by consensus algorithms. For instance, in the Proof of Work (PoW) consensus algorithm adopted by the Bitcoin network, nodes collectively solve a mathematical problem, with the algorithm dynamically adjusting the difficulty of the problem. The node that solves the problem first becomes the mining node and has a higher probability of being a mining node if it possesses greater computational power, leading to the receipt of block packaging rewards. In another consensus algorithm, Proof of Stake (PoS)\cite{King2012PPCoinPC}, nodes are ranked based on the quantity of stake they hold, with the node having the most stake becoming the mining node. Compared to PoW, PoS significantly reduces the computational resource consumption associated with competing for mining rights. However, it faces the challenge of centralization of packaging rights, as a single node may consecutively obtain packaging rights, posing a threat to consensus security.

\subsection{$\epsilon$-Diffeiential Privacy}

Differential privacy achieves privacy protection by introducing noise within a certain range into the training dataset or model parameters \cite{Dwork2014TheAF}, preventing attackers from inferring whether specific data items are present in the training data. Differential privacy was first proposed by Dwork in 2006 and requires trust between data owners and data administrators who send accurate \cite{Dwork2006DifferentialP}, unmodified data. Due to the absence of a trusted third party, data administrators uniformly add noise to the data before providing it to a third party for statistical queries.

We define an algorithm as $\epsilon$-differential private if it satisfies a certain degree of Central Differential Privacy. Here, $\epsilon \in \mathbb{R}^+$ represents privacy loss or privacy leakage, and a smaller $\epsilon$ indicates a higher level of privacy protection. Let $DS^n$ be the set of all datasets, $DS, DS' \in DS^n$, $DS \neq DS'$, and $DS$ has one entry different from $DS'$, meaning one entry in $DS$ can be removed to obtain $DS'$. The function $\digamma$ is a query function that maps datasets to real numbers: $\digamma: DS^n \rightarrow \mathbb{R}^d$. The sensitivity of the function $\digamma$ is defined as follows:

$$
\mathbb{S}_\digamma = \max \limits_{DS, DS'} ||\digamma(DS) - \digamma(DS') ||_1
$$
where $|| \cdot ||_1$ denotes the $l_1$ norm.


Hence, an algorithm $\Gamma$ is referred to as $(\epsilon, 0)$-differentially private if and only if, for $\forall DS, DS' \in D^n$, and $O \subseteq Y$, where $Y$ represents the set of all possible outputs, the following condition holds:

$$
Pr[\Gamma(DS) \in O] \leq exp(\epsilon)\cdot Pr[\Gamma(DS') \in O ]
$$

Here, $Pr$ denotes probability, and $Pr[\Gamma(DS) \in O]$ represents the probability that the output of $\Gamma$ applied to dataset $DS$ falls within a certain set $O$. Similarly, $Pr[\Gamma(DS') \in O]$ denotes the probability that the output of $\Gamma$ applied to dataset $DS'$ falls within the set $O$.
If $\digamma_1$ satisfies $\epsilon_1$-differential privacy, and $\digamma_2$ satisfies $\epsilon_2$-differential privacy, then we can get:

$\digamma_{1,2} = (\digamma_1, \digamma_2)$ satisfies ($\epsilon_1 + \epsilon_2$)-DP.

which is the \textit{composition theorem}.

%
%
%
%
%



\subsection{Mixing Methods}
%

The mixing mechanism was first proposed by D. Chaum in 1981 \cite{Chaum1981UntraceableEM}. In this approach, both communicating parties utilize an intermediary to transmit communication information that has undergone asymmetric encryption. Attackers are unable to discern the identities of the communicating parties or the connection between them through the analysis of encrypted information. Presently, mixing mechanisms can be classified into centralized and decentralized methods. In centralized mixing \cite{Bonneau2014MixcoinAF, Valenta2015BlindcoinBA,  Heilman2017TumbleBitAU}, users initially send tokens to a centralized mixing server. The central server aggregates users' transaction requests into a single transaction and decomposes the original output into outputs of equal value. Subsequently, the server proposes the transaction to the blockchain and charges a certain service fee. For example, if user $U_1$ wishes to initiate an anonymous transaction of $x_c$ tokens to user $U_2$, under the centralized mixing method, assuming the server charges a service fee of $fee_x$, $U_1$ first sends a transaction of $x_c + fee_x$ to the anonymous server. After receiving $U_1$'s transaction, the server deducts the server fee ($fee_x$), splits the transfer transaction of value $x_c$ into multiple transactions with different values, and publishes them on the blockchain network using its own address instead of the original address.

In decentralized mixing \cite{Ruffing2014CoinShufflePD, Schnoering2023HeuristicsFD }, various users collectively form an organization and use protocols for effective mixing. In this approach, if user $U_3$ wishes to initiate an anonymous transaction of $x_d$ tokens to user $U_4$, $U_3$ first seeks nodes willing to collaboratively create an anonymous transaction. $U_3$ forms an organization with these nodes, where some nodes assume the role of coordinators. Nodes in the organization send the amounts they want to use for the anonymous transaction to the coordinator. The coordinator aggregates similar transaction requests into a single transaction and decomposes the original transfer amount into multiple outputs of equal value. Then, nodes wishing to initiate the anonymous transaction confirm the creation of the transaction. The coordinator sends the anonymous transaction to the blockchain network, and nodes involved in the anonymous transaction collectively pay the split anonymous transaction fee.

\section{Privacy Protected Blockchain-based Federated Learning model}
\label{Privacy Protected Blockchain-based Federated Learning model}
In this section, we present the proposed model named PPBFL, which utilizes blockchain and adds differential noise to protect model parameters' privacy. In this process, PPBFL provides safety, high efficiency and low computation resource consumption.
\subsection{ Problem Statement}


%
%
%
We consider a scenario in a FL model where only one node in the local training nodes is known to be benign, while the others are untrusted nodes. There is a possibility that malicious nodes attempt to infer the local parameters of the benign node based on the global model, thereby obtaining information about the data distribution of the benign node. Assuming that the aggregation server is also untrusted and the communication network is susceptible to a man-in-the-middle attack, how can we protect the model parameters of the benign node and prevent inference attacks?
\begin{figure}[htbp]
	\centering
	\includegraphics[width=\columnwidth]{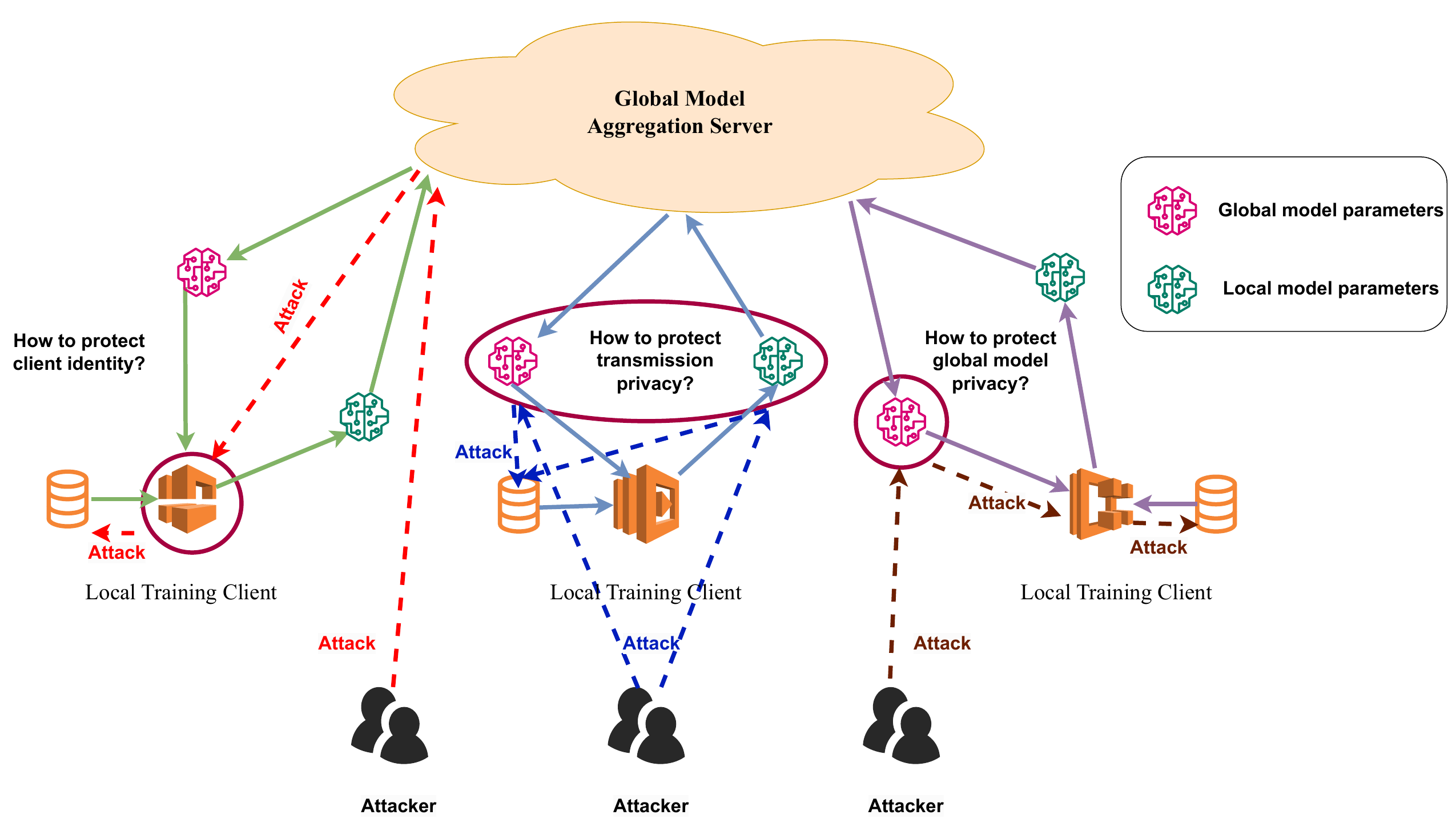}
	\caption{Problem Statement in PPBFL}
	\label{Problem Statement}
\end{figure} 
While this scenario is extreme, it reflects three critical privacy and security issues in current FL:
\begin{enumerate}[1. ]
	\item Privacy and security issues with the identity of local training nodes. In centralized FL, local training nodes need to send their local model parameters to the server. By exploiting the identity information of local training nodes, attackers may compromise the server to directly access the parameters of benign nodes, initiating inference attacks.
	\item Privacy and security issues during the transmission of model parameters. During transmission, attackers may intercept network traffic through network attacks, implementing man-in-the-middle attacks to obtain the local model parameters of benign nodes and infer the local data distribution.
	\item Privacy and security issues faced by the global model. Malicious nodes may collude to exclude their model parameters from the global model, obtaining specific local model parameters from the global model. This allows them to launch inference attacks on the local training data distribution of benign nodes.
\end{enumerate}

\subsection{  Model Overview}
\begin{figure*}[htbp]
	\centering
	\includegraphics[width=.8\textwidth]{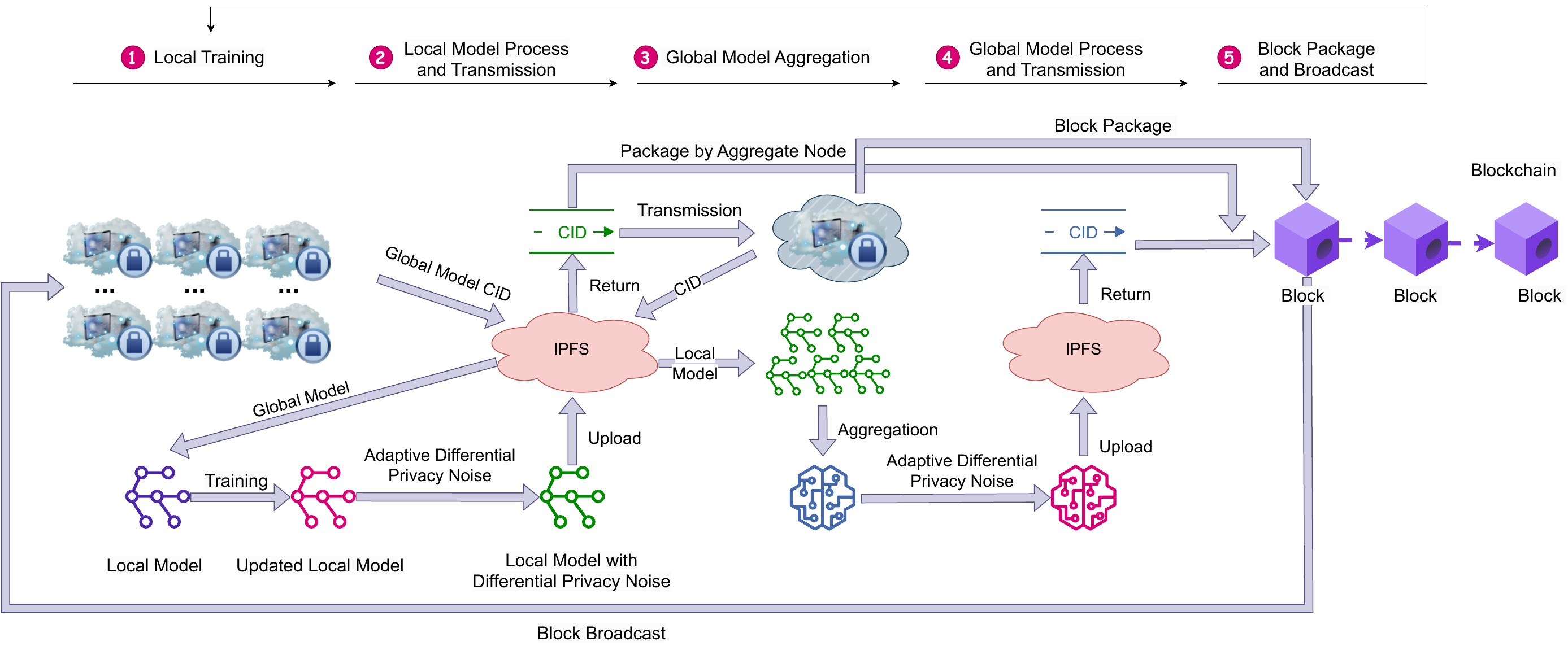}
	\caption{Model Structure in PPBFL}
	\label{Model Structure in PPBFL}
\end{figure*} 

Fig. \ref{Model Structure in PPBFL} shows an overview of our PPBFL model architecture.  PPBFL's design has the following goals:
%
%
%
%
\begin{enumerate}
	\item Generating a high-performance global model without revealing the identity of local training clients;
	\item Protecting local model parameters from disclosure during transmission;
	\item Ensuring the privacy of global model parameters to prevent malicious clients from inferring local model parameters based on the global model and subsequently inferring the data distribution of local training nodes;
\end{enumerate}

We propose PPBFL to achieve the above objectives by integrating FL with technologies such as blockchain, differential privacy, etc., to safeguard the privacy of FL training nodes and the transmission process.

Fig. \ref{Model Structure in PPBFL} shows PPBFL has following steps:
\begin{enumerate}
	\item Local training nodes training local model;
	\item Local training nodes add local differential privacy noise to local model, upload to IPFS, get CID, send to aggregation node;
	\item Aggregation node get local models from IPFS by CIDs, aggregate global model;
	\item Aggregation node upload global model to IPFS, get CID; 
	\item Aggregation node package transactions and global model CID into block, broadcast the new block.
\end{enumerate}

As illustrated in Fig. \ref{Model Structure in PPBFL}, in the PPBFL model, local training nodes in FL are interconnected through a blockchain network. After obtaining the Content Identifier (CID) of the global model from the blockchain, they download global model parameters from the InterPlanetary File System (IPFS) for local training. The updated local model is uploaded to IPFS and retrieved by the aggregation node for aggregation. The aggregated global model is then uploaded to IPFS, and the CID pointing to the location of the global model in IPFS is stored on the blockchain.

Storing local and global model parameters in IPFS returns a CID. The CID is essentially a hash based on the content of the stored file. Compared to the original file, the CID has a smaller volume. Storing all CIDs generated during the FL process on the blockchain does not excessively consume storage space. Additionally, it improves efficiency in network transmission during the process.


%
%

Both local model parameters and global model parameters undergo differential privacy processing before transmission. Applying differential privacy to local models protects the original parameters from disclosure, thereby ensuring the privacy of local data. Applying differential privacy to global models safeguards against inference attacks and protects the local data of benign nodes. According to the \textit{composition theorem} in differential privacy, when there are many local models in FL, the security guarantee of the combined global model will be reduced. To solve this problem, we apply reverse differential privacy to the global model to ensure effective model training.

The aggregation node finds the local model based on the CID stored in the blockchain. As the aggregation node is untrustworthy, we employ ring signature technology to protect the identity of local training nodes from being revealed. While blockchain technology enhances the traceability of transaction content and improves the security of FL algorithms, the transparency of blockchain content allows all nodes in the network to view it. Using ring signature hides the CID contained in transactions, providing greater efficiency compared to traditional ring signatures and addressing the issue of transparent data in blockchain technology.

Furthermore, we propose the Proof of Training Work consensus algorithm. Based on the training speed of local training nodes, we select the node with the fastest training speed as the aggregation node. The aggregation node receives additional rewards for aggregation and packaging, thereby incentivizing the active participation of local training nodes in FL tasks.

\subsection{ Initialization}
%
%
%
%
%
%
%

The definition of PPBFL is as follows:

$$
PPBFL = \{  \mathbb{AN}, \mathbb{LT}, \mathbb{BN}, \mathbb{GM}, \mathbb{LM}, PoTW, IPFS \}
$$

$\mathbb{AN} = \{AN_{\tau}^1, AN_{\tau}^2, \cdots, AN_{\tau}^\psi \}$ where $\psi \in \mathbb{Z}^+$ is the round number in FL, and $tau$ is the node ID,  represents the set of aggregation nodes responsible for aggregating the global model in FL. These nodes simultaneously have the packaging power for the current round of the blockchain. In PPBFL, $\mathbb{AN}$ is chosen by the consensus algorithm PoTW.

$\mathbb{LT} = \{LT_1^\psi, LT_2^\psi, \cdots, LT_{\alpha}^{\psi} \}$ represents the set of local training nodes in FL. $LT_{\alpha}^{\psi}$ represents the local training node with ID $\alpha$ in the $\psi$-th round.

$\mathbb{BN} = \{BN_1, BN_2, \ldots, BN_{\gamma}\}$ , $\gamma \in \mathbb{Z}^+$ is the node ID, $\mathbb{BN}$ represents the set of nodes that not participate in the FL task but joined the blockchain network and stores block, which is blockchain nodes. All FL nodes are part of the blockchain network and share a ledger as blockchain nodes. 

$\mathbb{GM} = \{GM_1^\chi, GM_2^\chi, \cdots, GM_\psi^\chi \}$ represents the set of global models. $GM_\psi^\chi $ represents the global model updated in the $\psi$-th round, and the node with ID $\chi$ is responsible for aggregating this global model.

$\mathbb{LM} = \{LM_\psi^1, LM_\psi^2, \cdots, LM_\psi^\eta\}$ represents the set of local models. $LM_\psi^\eta$ represents the local model updated by the node with ID $\eta$ in the $\psi$-th round.

$PoTW$ represents the Proof of Training Work consensus algorithm. In PPBFL, we propose PoTW to select nodes for aggregating the global model and to package transactions into the blockchain during this period.

$IPFS$ represents the InterPlanetary File System. In PPBFL, we use IPFS to store $\mathbb{LM}$ and $\mathbb{GM}$. The Content Identifier (CID) returned by IPFS is stored in the blockchain, reducing the storage resource consumption of the blockchain and improving network transmission speed.

\subsection{Local Training and Model Communication }

$\mathbb{LT}$ in PPBFL will begin local training after getting the $\mathbb{GM}$ from IPFS with the CID in blockchain. $\mathbb{LT}$ may have various devices, such as phones, computers, autonomous vehicles, routers, televisions and so on, they can train the same model with their local data, with their computation power. 
%


IPFS helps reduce blockchain storage consumption. Blockchain provides transaction traceability, which can assist in tracing FL records. However, every block is stored on each node. Storing blocks requires local storage space on nodes, and due to variations in local training node types, storage capacities also differ. FL typically involves multiple communication rounds, and there may be numerous $\mathbb{LT}$. If both $\mathbb{LM}$ and $\mathbb{GM}$ are stored in the blockchain, it could impose significant storage pressure on nodes.

Storing $\mathbb{LM}$ and $\mathbb{GM}$ parameters in IPFS, with IPFS returning Content Identifiers (CIDs) as hash numbers after file hashing operations, occupies less space. Storing model parameters in IPFS and transmitting CIDs through the blockchain can significantly improve network transmission efficiency and reduce storage pressure on nodes.

Before $\mathbb{LM}$ parameters start to transmit, we employ two methods to protect model parameter privacy and node identity privacy, namely, ring signature-based transactions and dual adaptive differential privacy mechanism.

\subsection{ Adaptive Local Differential Privacy Mechanism}


We ensure the privacy of $\mathbb{LM}$ parameters by adding differential privacy noise to the $\mathbb{LM}$. According to the composition theory, in order to prevent a decrease in $\mathbb{GM}$ security caused by a large number of $\mathbb{LM}$ aggregation in FL, both $\mathbb{LT}$ and $\mathbb{AN}$ in the PPBFL add differential privacy noise in opposite directions to the $\mathbb{LM}$ and $\mathbb{GM}$, respectively.
\begin{figure}[htbp]
	\centering
	\includegraphics[width=\columnwidth]{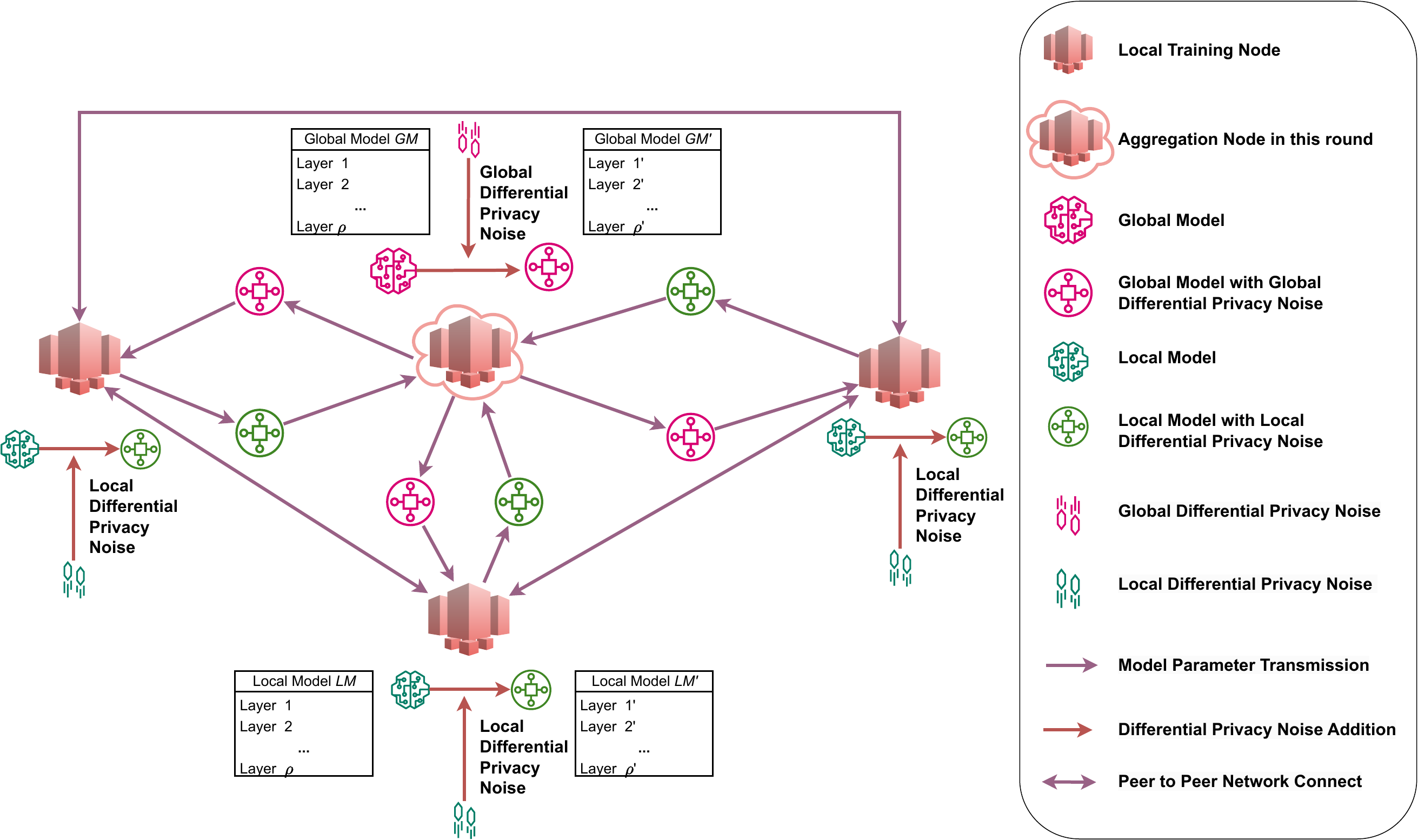}
	\caption{Dual Adaptive Local Differential Privacy Mechanism in PPBFL}
	\label{Dual Adaptive Local Differential Privacy Mechanism}
\end{figure} 
In PPBFL, considering the trade-off between privacy protection effectiveness and efficiency, we employ $\epsilon$-DP, providing a certain level of flexibility to the protected $\mathbb{LM}$ parameters to enhance the efficiency of differential privacy. Building upon $\epsilon$-DP, we introduce an adaptive differential privacy algorithm to address the issue of uncontrollable noise levels in $\epsilon$-DP.  Subsequently, we compute the distance between the parameters of the local model ($\mathbb{LM}$) and the central point of weights pertaining to the respective layer in the preceding round's $\mathbb{GM}$, adjusting for differential privacy noise.

The calculation formula for the center value of weights in  layer $\rho$ of the global model is presented as shown in Equation \ref{C_g}. We obtain the center point by averaging the maximum and minimum values of weights within the same layer.

\begin{equation}
	\mathbb{C}_g^{\rho}= \frac{max_g^{\rho}+ min_g^{\rho}}{2}
\label{C_g}
\end{equation}

\begin{figure*}
\begin{equation}
\zeta_{l;\xi}^{\psi;\rho} =  \left\{
\begin{array}{ll}
	|max_l^{\rho}- \mathbb{C}_g^{\rho}|, & \text{if} (|max_l^{\rho}- \mathbb{C}_g^{\rho}|  > | \mathbb{C}_g^{\rho}-  min_l^{\rho}|),\\
	|\mathbb{C}_g^{\rho}-  min_l^{\rho}|, & \text{if} (|\mathbb{C}_g^{\rho}-  min_l^{\rho}| < |max_l^{\rho}- \mathbb{C}_g^{\rho}|).
\end{array}
\right.
\label{radius_l}
\end{equation}
\end{figure*}

In equation \ref{radius_l}, $\zeta_{l;\xi}^{\psi;\rho}$ represents the radius of the model parameters in layer $\rho$ of  $LT_\xi$ during the $\psi$-th round, with $\mathbb{C}_g^{\rho}$ as the center.

Thus, the model weights for the layer are all within the interval $[\mathbb{C}_g^{\rho}- \zeta_{l;\xi}^{\psi;\rho}, \mathbb{C}_g^{\rho}+ \zeta_{l;\xi}^{\psi;\rho}]$. For each layer in the $\mathbb{LM}$ of each client, calculate the distance between each weight and $	\mathbb{C}_g^{\rho}$:

\begin{equation}
	\Delta = \omega - \mathbb{C}_g^{\rho}
\end{equation}
Hence, $\omega$ is the weight of $\mathbb{LM}$, $\Delta$ is the distance from the weight to $\mathbb{C}_g^{\rho}$.


%
We utilize the Bernoulli distribution function to select a random number $m \in \{0,1\}$. 
\begin{equation}
	Pr_{\mathbb{L}}[m=1] = \frac{e^{\epsilon}}{2e^{\epsilon}+1}
\end{equation}
\begin{equation}
	Pr_{\mathbb{L}}[m=0] = 1 - Pr_{\mathbb{L}}[m=1] = \frac{e^{\epsilon}+1}{2e^{\epsilon}+1}
\end{equation}

Building upon the work presented in the paper \cite{Sun2020LDPFLPP,Miao2022CompressedFL} our proposed probability mass function (PMF) is designed to be closer to 1/2.

\begin{equation}
P(\omega) =  \left\{
\begin{array}{ll}
	\mathbb{C}_g^{\rho}+ \Delta \frac{e^\epsilon +1}{e^\epsilon }, & \text{if}\  m=1, \\
	\mathbb{C}_g^{\rho}+ \Delta  \frac{e^\epsilon }{e^\epsilon + 1}, & \text{if}\ m=0.
\end{array}
\right.
\label{POMEGA}
\end{equation}
%
%

Therefore, $\Delta$ represents the distance between weight and the center point of the $\mathbb{GM}$. 

The existing method presented in \cite{Sun2020LDPFLPP, Miao2022CompressedFL}, introduces local adaptive differential privacy. This approach calculates the range of weights of the $\mathbb{LM}$ in different layers, adapting the addition of differential privacy. Building upon this, we calculate the value range of the weights of the previous round's $\mathbb{GM}$ in different layers. This approach avoids the uncertainty issue associated with the value range of individual $\mathbb{LM}$ and ensures that different $\mathbb{LM}$ add differential privacy noise within the same interval. This resolves the problem of varying differential privacy noise ranges across different $\mathbb{LM}$. Compared to computing the comparison standard by comparing the weight with the center value of the $\mathbb{LM}$ parameters, our approach, which utilizes the calculated difference to determine the weighted values after adding differential noise, standardizes the comparison criteria for different $\mathbb{LT}$ in the same round.

\begin{algorithm}
	\renewcommand{\algorithmicrequire}{\textbf{Require:}}
	\caption{Local model differential privacy}
	\label{alg:PPBFL Model}	
	\begin{algorithmic}[1]
		\REQUIRE{Local Training Data}
		\output{Global Model}
		\STATE	$\mathbb{C}_g^{\rho}= \frac{max_g^{\rho}+ min_g^{\rho}}{2}$ ; \COMMENT{calculate each layer's center of last round global model} 
		\FOR{each $LT$ in $\mathbb{LT}$}
		
		\FOR{each weight $\omega$ in LT.LM}
		\STATE $\Delta = \omega - \mathbb{C}_g^{\rho}$; \COMMENT{calculate the difference between $\omega$ and $\mathbb{C}_g^{\rho}$} \;
		
		\STATE	m $\leftarrow$ random.choice(\{1,0\}) with probability distribution  \{$\frac{e^{\epsilon}}{2e^{\epsilon}+1}$, $\frac{e^{\epsilon}+1}{2e^{\epsilon}+1} $\} \;
		
		\IF{m = 1 }
		\STATE $ \omega^{'} \leftarrow \mathbb{C}_g^{\rho}+ \Delta \frac{e^\epsilon +1}{e^\epsilon } $ \;
		
		\ELSIF{m = 0}
		\STATE 	$\omega^{'} \leftarrow \mathbb{C}_g^{\rho}+ \Delta \frac{e^\epsilon }{e^\epsilon + 1}$ \;
		\ENDIF
		\ENDFOR
		\ENDFOR

	\end{algorithmic}
\end{algorithm}

\subsection{Ring Signature-based Mixing CID Mechanism}

In PPBFL, we inspired by the ring signature technique within Ring Confidential Transactions (RingCT) for mixing transactions. We propose a mixing CID mechanism to hide the node address initiating the transaction, achieving transaction anonymity in PPBFL.



In \cite{Noether2015RingSC}, the authors proposed a mixing mechanism applied in Bitcoin and Monero. In this paper, building upon the centralized mixing mechanism, we introduce a mixing CID mechanism tailored for FL, extending the identity anonymity of transaction initiators in PPBFL based on ring signatures.

Our proposed mixing CID mechanism allows nodes, when initiating transactions, to independently include content from other transactions within the same time period. After uploading their $\mathbb{LM}$ parameters to IPFS, $\mathbb{LT}$ select one or more CIDs included in transactions received during the current round. They then combine their model's CID with those obtained from other transactions and package them into a blockchain transaction, broadcasting it across the blockchain network. Upon receiving the transaction, aggregate nodes compare the CIDs contained in the newly received transaction with those from previously received transactions to determine if the CIDs have already been verified. For verified CIDs, aggregate nodes refrain from redundant retrieval of $\mathbb{LM}$. In case the CID is unverified, the aggregate node sends a request to IPFS to obtain the $\mathbb{LM}$ parameters based on the CID. A single transaction may include multiple CIDs, and aggregate nodes validate each CID individually.

\subsection{ Proof of Training Work}

The consensus algorithm ensures the addition of correct blocks to the blockchain. The primary goal is to ensure consensus among various nodes in the distributed system regarding the system state or transactions. Blockchain, as a distributed system, employs consensus algorithms to address issues arising from network delays, node failures, or malicious activities, ensuring the consistency, reliability, and security of the system.
In our approach, we build upon the PoW consensus algorithm but replace the mathematical problem-solving component with $\mathbb{LM}$ training in FL. In PoW, solving mathematical problems serves no practical purpose; it is merely for competing for packaging power, leading to significant waste of energy and computational resources. The Proof of Training Work (PoTW) consensus algorithm retains the selection of nodes as packaging nodes based on the proof of their training efforts, which are demonstrated through the speed of model training. Assuming a fixed number of training rounds, which cannot be reduced to shorten the training time, we select the local training node that spends the least time training the same model. This node becomes the $\mathbb{GM}$ aggregation node in the next round of FL, also possessing the power to package blocks and receive packaging rewards. A shorter training time for the same model indicates that the node has invested more computational resources in the FL training task. Compared to PoW, the PoTW consensus algorithm maintains security and decentralization characteristics while transforming the consumed computing resources into the task of federated model training, addressing the issues of energy and resource consumption associated with PoW. In contrast to PoS, the PoTW consensus algorithm improves the security and decentralization capabilities of the consensus algorithm without altering the local training time for FL tasks.

\subsection{ Global Model Aggregation and Reverse Differential Privacy}
In FL, if $\mathbb{LT}$ introduce differential privacy (DP) noise to their $\mathbb{LM}$, assuming there are $\vartheta$ nodes, and each node adds $\epsilon$-DP noise to its $\mathbb{LM}$, according to the \textit{composition theorem}, the aggregated global model satisfies ($\sum_{i=1}^{\vartheta} \epsilon_i$)-DP. However, after aggregation, the privacy assurance of the global model is weakened by the influence of $\mathbb{LT}$. To address this, we propose the addition of adaptive differential privacy noise to the global model, counteracting the differential privacy noise added in $\mathbb{LM}$. This approach aims to enhance the security of the global model.
After the selection of $\mathbb{AN}$ by PoTW consensus algorithm, $\mathbb{AN}$ collects transactions within the current time period. Based on the Content Identifier (CID) contained in the transactions,  $\mathbb{AN}$  retrieves $\mathbb{LM}$ stored in the IPFS. We utilize the FedAvg algorithm to aggregate the $\mathbb{LM}$, obtaining the global model for the current round. As differential privacy noise exists in various $\mathbb{LM}$, to avoid the security guarantee reduction problem caused by too many $\mathbb{LM}$ in FL based on \textit{composition theorem}, PPBFL introduces differential privacy noise in the opposite direction to the $\mathbb{GM}$. This approach ensures mutual cancellation of the differential privacy noise. Additionally, the $\mathbb{GM}$ may compromise the privacy of $\mathbb{LM}$ of benign nodes. By adding differential privacy noise to the $\mathbb{GM}$ in PPBFL, the privacy of $\mathbb{GM}$ parameters is safeguarded, preventing them from being exploited in inference attacks. Despite the presence of differential privacy noise in multiple $\mathbb{LM}$, the FedAvg aggregation algorithm averages $\mathbb{LM}$ parameters, allowing us to add differential privacy noise to the $\mathbb{GM}$ only once.
$$
\omega_{\psi+1}^g = \sum\limits_{k=1}^{\vartheta} \omega_{\psi}^k
$$

In the context of $\mathbb{GM}$ differential privacy, we calculate the distance between the parameters of the $\mathbb{GM}$ and its own center point for the current round. We then introduce differential privacy noise based on this distance. As the $\mathbb{GM}$ for the current round has been updated compared to the previous round, we use the current round's $\mathbb{GM}$ to calculate the center points for each layer. This ensures the addition of differential privacy noise to maintain the privacy of the $\mathbb{GM}$ in the current round.

\begin{equation}
Pr_{\mathbb{G}}[m=1] = \frac{e^{\epsilon}}{2e^{\epsilon}+1}
\label{PRG1}
\end{equation}
 
 \begin{equation}
Pr_{\mathbb{G}}[m=0] = 1 - Pr_{\mathbb{G}}[m=1] = \frac{e^{\epsilon}+1}{2e^{\epsilon}+1} 
\label{PRG0}
\end{equation}


\begin{equation}
G(\omega) =  \left\{
\begin{array}{ll}
	\mathbb{C}_g^{\rho}+ \Delta  \frac{e^\epsilon -1}{e^\epsilon}, & \text{if}\ m=1, \\
	\mathbb{C}_g^{\rho}+ \Delta \frac{e^\epsilon + 2}{e^\epsilon + 1}, & \text{if}\ m=0.
\end{array}
\right.
\label{GOMEGA}
\end{equation}
Equations \ref{PRG1} and \ref{PRG0} respectively denote the probabilities of the Bernoulli variable $m$ being 1 and 0 in the context of differential privacy in the global model. Equation \ref{GOMEGA} represents the adaptive differential privacy algorithm for the global model that we propose, accounting for the cases when $m$ takes on values of 1 and 0.

\subsection{Block Package and Global Model Communication}
%

After aggregating the $\mathbb{GM}$, the $\mathbb{AN}$ uploads it to IPFS. IPFS returns the CID of the $\mathbb{GM}$. With the authority to package blocks, the $\mathbb{AN}$ initiates a transaction by including the CID of the $\mathbb{GM}$. This transaction is then bundled with other transactions containing $\mathbb{LM}$ CIDs received during the same period and added to the blockchain.

\section{ Security Analysis}
\label{ Security Analysis}



In this section, we will analyze the privacy guarantees and utility evaluation of PPBFL. While FL protects local data privacy by sharing model parameters, potential security risks and threats to the privacy of client-side local data still exist.

One potential attack involves eavesdropping on the transmission of $\mathbb{LM}$ to obtain their parameters, enabling model inference attacks. Another method involves leveraging the publicly accessible $\mathbb{GM}$ on the network, excluding self and colluding $\mathbb{LM}$ parameters to deduce specific benign node $\mathbb{LM}$ parameters for inference attacks. Both attack methods aim to deduce the local data distribution of benign nodes from their $\mathbb{LM}$ parameters. Based on the post-processing characteristics of differential privacy, the addition of noise to $\mathbb{LM}$ and $\mathbb{GM}$ ensures that, if differential privacy is satisfied, no model privacy is leaked. For untrusted servers and $\mathbb{LT}$ in FL, if nodes add differential privacy noise before transmitting data, even if parameters are acquired by malicious nodes, local data will not be compromised. 

In the following, we provide a rigorous privacy proof for the dual adaptive differential privacy mechanism, which encompasses both local model differential privacy and global model differential privacy.
\subsection{ Local Model Differential Privacy}


Theorem 1: For any weight $w \in [\mathbb{C}_g - \zeta_{l;\xi}^{\psi;\rho}, \mathbb{C}_g + \zeta_{l;\xi}^{\psi;\rho}]$, the proposed mechanism $P(\omega)$ in Equation \ref{POMEGA}, which encompasses both $\mathbb{LM}$ differential privacy and $\mathbb{GM}$ differential privacy, satisfies $\epsilon-DP$.

Proof. The range of $\omega$ in layer $\rho$ of local model $LM_\xi$ is $[\mathbb{C}_g^{\rho}-\zeta_{l;\xi}^{\psi;\rho}, \mathbb{C}_g^{\rho}+\zeta_{l;\xi}^{\psi;\rho}]$. If $\omega^* = \mathbb{C}_g^{\rho}+ \Delta \frac{e^\epsilon +1}{e^\epsilon }$, then for any $\omega, \omega^{\prime} \in [\mathbb{C}_g^{\rho}-\zeta_{l;\xi}^{\psi;\rho}, \mathbb{C}_g^{\rho}+\zeta_{l;\xi}^{\psi;\rho}]$.

\begin{equation}
	\begin{aligned}
\frac{Pr[P(\omega) = \omega^*]}{Pr[P(\omega^{\prime}) = \omega^*]} & \leq \frac{\mathop{max}\limits_{\omega} Pr[P(\omega) = \omega^*]}{\mathop{min}\limits_{\omega^{\prime}} Pr[P(\omega^{\prime}) = \omega^*]}\\
&= (\frac{e^{\epsilon}}{2e^{\epsilon}+1}) / (\frac{e^{\epsilon}+1}{2e^{\epsilon}+1})
= \frac{e^{\epsilon} }{e^{\epsilon} +1}\\
	\end{aligned}
\end{equation}

According to the definition of $\epsilon$-differential privacy, when $ \frac{e^{\epsilon} }{e^{\epsilon} +1} < e^{\epsilon}$, $P(\omega)$ satisfies $\epsilon$-differential privacy. So $P(\omega)$ satisfies $\epsilon$-differential privacy.

Lemma 1: The dual adaptive differential privacy algorithm introduces a bias of 0 when calculating the average weight, i.e., $\mathbb{E}[\overline{P(\omega)}] = \overline{\omega}$.

Proof. For any weight $\omega$ from any client $\mathbb{LT}$,

\begin{equation}
	\begin{aligned}
\mathbb{E}[P(\omega)]
& = (\mathbb{C}_g^{\rho}+ \Delta \frac{e^\epsilon +1}{e^\epsilon }) \cdot \frac{e^{\epsilon}}{2e^{\epsilon}+1} \\
&+ (\mathbb{C}_g^{\rho}+ \Delta  \frac{e^\epsilon }{e^\epsilon + 1}) \cdot \frac{e^{\epsilon}+1}{2e^{\epsilon}+1}\\
&= \frac{2\mathbb{C}_g^{\rho}e^{\epsilon}+\mathbb{C}_g^{\rho}}{2e^{\epsilon}+1} + \frac{\Delta(2e^{\epsilon}+1)}{2e^{\epsilon}+1}\\
&=\mathbb{C}_g^{\rho}+\Delta \\ 
&= \omega\\
	\end{aligned}
\end{equation}

\begin{equation}
	\begin{aligned}
\mathbb{E}[\overline{P(\omega)}] 
&= \mathbb{E}[\frac{1}{N}\sum\limits_{n=1}^{N}P(\omega)] = \frac{1}{N} \sum\limits_{n=1}^{N} \mathbb{E}[P(\omega)]\\
& = \frac{1}{N} \sum\limits_{n=1}^{N} \omega \\ &= \overline{\omega}\\
\end{aligned}
\end{equation}

Lemma 2. Let $P$ be the proposed data perturbation algorithm. Given any weight $\omega$, the variance of the algorithm is $((\zeta_{l;\xi}^{\psi;\rho})^2)/(e^{\epsilon}(e^{\epsilon}+1))$.

Proof. The variance of $\omega$ with noise is 

\begin{equation}
	\begin{aligned}
&Var[P(\omega)] = \mathbb{E}(P^2(\omega))- \mathbb{E}^2(P(\omega))\\
& = Var[\mathbb{C}_g^{\rho}+\Delta^*] \\
& = Var[\Delta^*] = \mathbb{E}(\Delta^{*2}) - \Delta^2\\
&= (\Delta \frac{e^\epsilon +1}{e^\epsilon })^2 \cdot  \frac{e^{\epsilon}}{2e^{\epsilon}+1} \\
& + (\Delta  \frac{e^\epsilon }{e^\epsilon + 1})^2 \cdot \frac{e^{\epsilon}+1}{2e^{\epsilon}+1} - \Delta^2 \\
&= \frac{\Delta^2}{e^{\epsilon}(e^{\epsilon} +1)}
\leq \frac{(\zeta_{l;\xi}^{\psi;\rho})^2}{e^{\epsilon}(e^{\epsilon} +1)}\\
\end{aligned}
\end{equation}

Lemma 3. Consider the estimated average weight represented by $\overline{P(\omega)}$, both the lower and upper bounds for this estimated average weight is: $ 0 \leq Var[\overline{P(\omega)}] \leq \frac{(\zeta_{l;\xi}^{\psi;\rho})^2}{N \cdot e^{\epsilon}(e^{\epsilon} +1)}$.

Proof. The variance of the estimated average weight is:

\begin{equation}
Var[\overline{P(\omega)}]  = Var[ \frac{1}{N} \sum\limits_{n=1}^{N} P(\omega)] 
= \frac{\Delta^2}{N \cdot e^{\epsilon}(e^{\epsilon} +1)}
\end{equation}


The range of $\Delta$ is [0, $\zeta_{l;\xi}^{\psi;\rho}$]. Therefore, substituting the maximum and minimum values of $\Delta$ into $Var[\overline{P(\omega)}]$, we obtain:

\begin{equation}
0 \leq Var[\overline{P(\omega)}] \leq \frac{(\zeta_{l;\xi}^{\psi;\rho})^2}{N \cdot e^{\epsilon}(e^{\epsilon} +1)}
\end{equation}

\subsection{Global Model Differential Privacy}

Theorem 2. Given any weight  $w \in [\mathbb{C}_g - \zeta_{g}^{\psi;\rho}, \mathbb{C}_g + \zeta_{g}^{\psi;\rho}]$ , where $\mathbb{C}_g^\rho$ is the center of $gm_\psi$'s range in layer $\rho$, $\zeta_{g}^{\psi;\rho}$  is the radius of model weight in layer $\rho$ of the global model $gm_\psi$, the proposed mechanism $P(\omega)$ in Equation $G(\omega)$ satisfies $\epsilon-DP$.

Proof. The range of $\omega$ in layer $\rho$ of global model  is $[\mathbb{C}_g^{\rho}-\zeta_{g}^{\psi;\rho}, \mathbb{C}_g^{\rho}+\zeta_{g}^{\psi;\rho}]$. If $\omega^* = \mathbb{C}_g^{\rho}+ \Delta \frac{e^\epsilon -1}{e^\epsilon }$, then for any $\omega, \omega^{\prime} \in [\mathbb{C}_g^{\rho}-\zeta_{g}^{\psi;\rho}, \mathbb{C}_g^{\rho}+\zeta_{g}^{\psi;\rho}]$.

\begin{equation}
\frac{\mathop{max}\limits_{\omega}Pr[G(\omega) = \omega^*]}{\mathop{min}\limits_{\omega^{\prime}}Pr[G(\omega^{\prime}) = \omega^*]} 
= \frac{e^{\epsilon}}{e^{\epsilon}+1}
\end{equation}

We can get the $G(\omega)$ satisfies $\epsilon-DP$.


Lemma 4. Algorithm $G(\omega)$ introduces a zero bias when calculating the average weight, i.e., $\mathbb{E}[\overline{P(\omega)}] = \overline{\omega}$.
$$
\mathbb{E}[G(\omega)] = \mathbb{C}_g^{\rho} + \Delta = \omega
$$

Lemma 5. Let $P$ be the proposed data perturbation algorithm, Given any weight $\omega$, the variance of the algorithm is $\frac{(\zeta_{l;\xi}^{\psi;\rho})^2}{e^{\epsilon}(e^{\epsilon}+1)}$.

Proof. The variance of $\omega$ with noise is

\begin{equation}
	\begin{aligned}
		Var[G(\omega)] 
		&= \mathbb{E}(G^2(\omega))- \mathbb{E}^2(G(\omega)) \\
		&= \frac{\Delta^2}{e^{\epsilon}(e^{\epsilon} +1)}
		\leq \frac{(\zeta_{l;\xi}^{\psi;\rho})^2}{e^{\epsilon}(e^{\epsilon} +1)}\\
	\end{aligned}
\end{equation}


\subsection{ Dual Differential Privacy}
Theorem 3. Algorithms $P(\omega)$ and $G(\omega)$ introduce zero bias when noise is simultaneously added.

Proof. For any weight $\omega$ from any local model $lt$ and in the $\mathbb{GM}$, when adding differential privacy noise simultaneously in both the $\mathbb{LM}$ and $\mathbb{GM}$,

\begin{equation}
	\begin{aligned}
&\mathbb{E}[P(\omega),G(\omega)]\\ 
&= \frac{\mathbb{E}[P(\omega)] + \mathbb{E}[G(\omega)] }{2}\\ 
& = ((\mathbb{C}_g^{\rho}+ \Delta \frac{e^\epsilon +1}{e^\epsilon }) \cdot (\frac{e^{\epsilon}}{2e^{\epsilon}+1})\\
&+ (\mathbb{C}_g^{\rho}+ \Delta  \frac{e^\epsilon -1}{e^\epsilon}) \cdot (\frac{e^{\epsilon}}{2e^{\epsilon}+1})\\
&+	(\mathbb{C}_g^{\rho}+ \Delta  \frac{e^\epsilon }{e^\epsilon + 1}) \cdot (\frac{e^{\epsilon}+1}{2e^{\epsilon}+1}) \\
&+ (\mathbb{C}_g^{\rho}+ \Delta \frac{e^\epsilon + 2}{e^\epsilon + 1}) \cdot (\frac{e^{\epsilon}+1}{2e^{\epsilon}+1})
	)/2\\
&= \mathbb{C}_g^{\rho}+ \Delta\\
\end{aligned}
\end{equation}

\section{ Experimental Results}
\label{Experimental Results}

In this section, we will delineate the experimental setup and evaluation criteria employed in the current study. Subsequently, we will assess and analyze the effectiveness of our differential privacy method, shedding light on its overall performance.

\subsection{Datasets}

In our experiments, we used MNIST and Fashion-MNIST as datasets, both of which are open source dataset. For each dataset, we examined the accuracy in   independent and identically distributed (IID) and not identically and independently distributed (Non-IID). Specially, we first test the accuracy of FL task with different $\epsilon$ ,  $\epsilon = \{0.5, 1, 2, 3, 4, 5$ ,  and then we test the accuracy not add global differential privacy noise. Moreover, we also test the performance while using  PoS and PoTW consensus, and compared the accuracy and device stakes in independent FL tasks.

\begin{itemize}
	
	\item MNIST: The MNIST dataset is composed of handwritten digits collected from 250 distinct individuals, resulting in a total of 70,000 images. The training set encompasses 60,000 images, while the test set comprises 10,000 images. All images are in grayscale and possess dimensions of 28×28 pixels, each showcasing a single handwritten digit.
	
	\item Fashion-MNIST: The Fashion-MNIST dataset encompasses 70,000 frontal images featuring a diverse array of fashion products distributed across 10 categories. The dataset mirrors the structure of the MNIST dataset in terms of size, format, and the division into training and test sets. It follows a 60,000/10,000 split for training and testing, with each image sized at 28×28 pixels and presented in grayscale.
\end{itemize}

\subsection{Model Structure}

For MNIST , we used two convolution layers, kernel size is 5 × 5,  maximum pooling layer is followed with each of them, and used two fully connected layer. The output channel of the first convolution layer are 32, and the second are 64. In the output layer, the output of fully connection layer is processed by the $ReLU$ function. 

For Fashion-MNIST,  we also used two convolution layers, both of them kernel size is 3 × 3,  maximum pooling layer is followed with each of them, and used three fully connected layer, one dropout layer followed by the first fully connected layer. The output channel of the first convolution layer are 32, and the second are 64, same as the channel in MNIST. In the output layer, the output of fully connection layer is also processed by the $ReLU$ function. 

\subsection{Runtime Environment}

Our experiments were conducted on a server, using PyTorch version 2.0.0, CUDA version 11.8, with a system equipped with 56GB of RAM and a 30GB hard disk, the GPUs in the server are  Tesla T4 (with a VRAM of 16GB).

\subsection{Accuracy Across Various $\epsilon$ Values in PPBFL}

\subsubsection{Accuracy of CAFL}

\begin{figure}[htbp]
	\centering
	\begin{subfigure}[b]{0.47\textwidth}
		\centering
		\includegraphics[width=\textwidth]{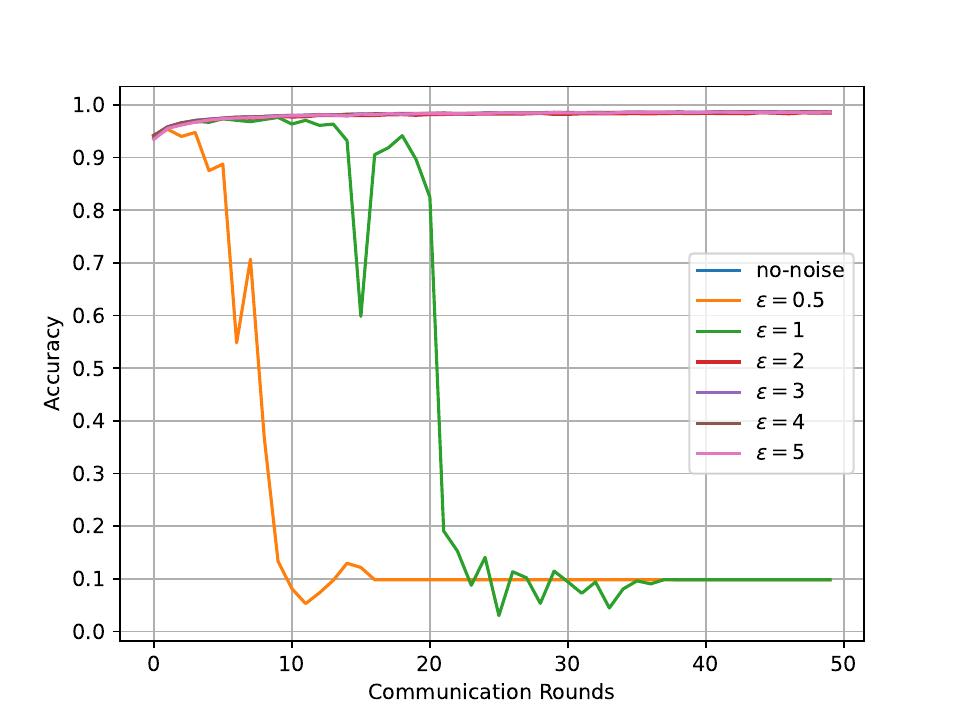}
		\caption{MNIST-IID}
		\label{CAFL-MNIST-IID}
	\end{subfigure}
	\begin{subfigure}[b]{0.47\textwidth}
		\centering
		\includegraphics[width=\textwidth]{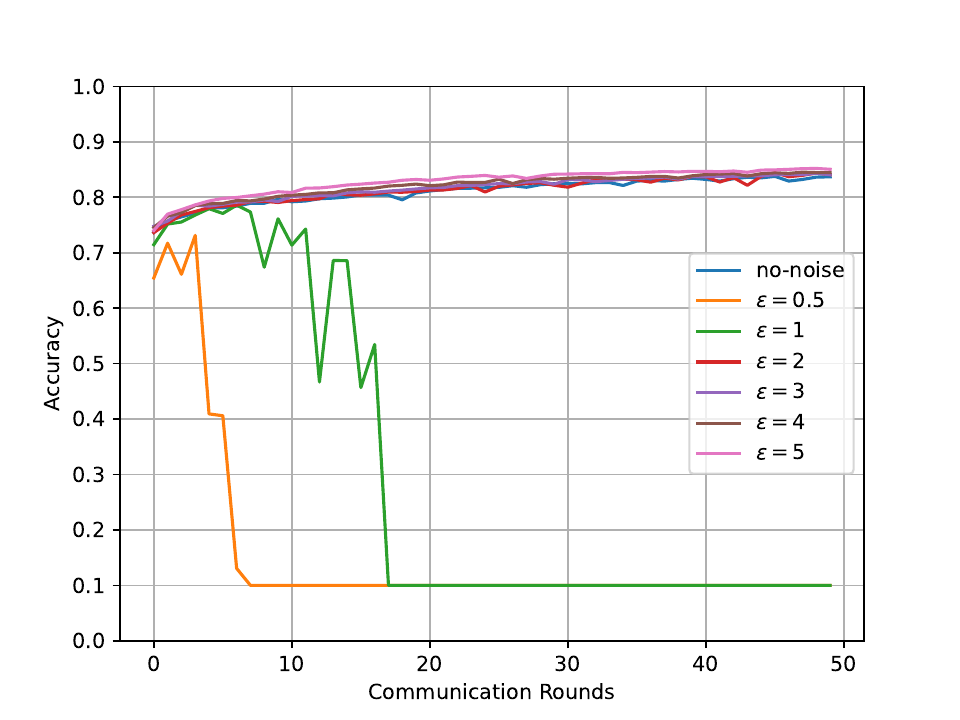}
		\caption{Fashion-MNIST-IID}
		\label{CAFL-Fashion-MNIST-IID}
	\end{subfigure} 
	\caption{Accuracy of CAFL under Different $\epsilon$ for MNIST and Fashion-MNIST task, $\epsilon = \{0.5, 1, 2, 3, 4, 5\}$ }
	\label{Accuracy of CAFL}
\end{figure}

In this study, we evaluate the performance of CAFL \cite{Miao2022CompressedFL}. We replace the distance of weights relative to the central point in the $\mathbb{GM}$ for the corresponding layer with the offset $\mu$ relative to the central point of the layer in its local model. Simultaneously, we set the perturbation equation to Equation \ref{CAFLM} \cite{Miao2022CompressedFL}, as illustrated in Figure \ref{Accuracy of CAFL}. The data follows an independent and identically distributed (IID) distribution.

\begin{equation}
	M(\omega) = w^* = \left\{
	\begin{array}{ll}
		c_l +\mu \cdot \frac{e^\epsilon +1}{e^\epsilon -1}, \textit{if}\  b = 1;\\
		c_l +\mu \cdot \frac{e^\epsilon -1}{e^\epsilon +1}, \textit{if} \ b = 0;\\
	\end{array}
	\right.
	\label{CAFLM}
\end{equation}
 the probability of the Bernoulli variable $b$ taking 1 of CAFL is shown in Equation \ref{CAFLPR}.
 \begin{equation}
 	Pr[b=1] = \frac{e^\epsilon -1 }{2e^\epsilon}
 	\label{CAFLPR}
 \end{equation}
 
 
 From Figure \ref{CAFL-MNIST-IID}, when adopting CAFL on the MNIST dataset with independently and identically distributed data, we observe that with $\epsilon$ set to 0.5, the model's accuracy decreases to 0.1 within the first 10 rounds. When $\epsilon$ is set to 1, the model's accuracy drops to around 0.1 just after 20 rounds, fluctuating for a while before stabilizing at 0.1. In Figure \ref{CAFL-Fashion-MNIST-IID}, for both $\epsilon$ values of 0.5 and 1, the model's accuracy on the Fashion-MNIST dataset decreases more rapidly to 0.1 compared to the MNIST dataset and remains stable.  In both the MNIST and Fashion-MNIST datasets, when $\epsilon$ exceeds 1, the model's accuracy remains consistent or similar to when no differential privacy noise is added.
 
\subsubsection{Accuracy of PPBFL with No Global Model Differential Privacy }

	\begin{figure}[htbp]
		\centering
		\begin{subfigure}[b]{0.47\textwidth}
			\centering
			\includegraphics[width=\textwidth]{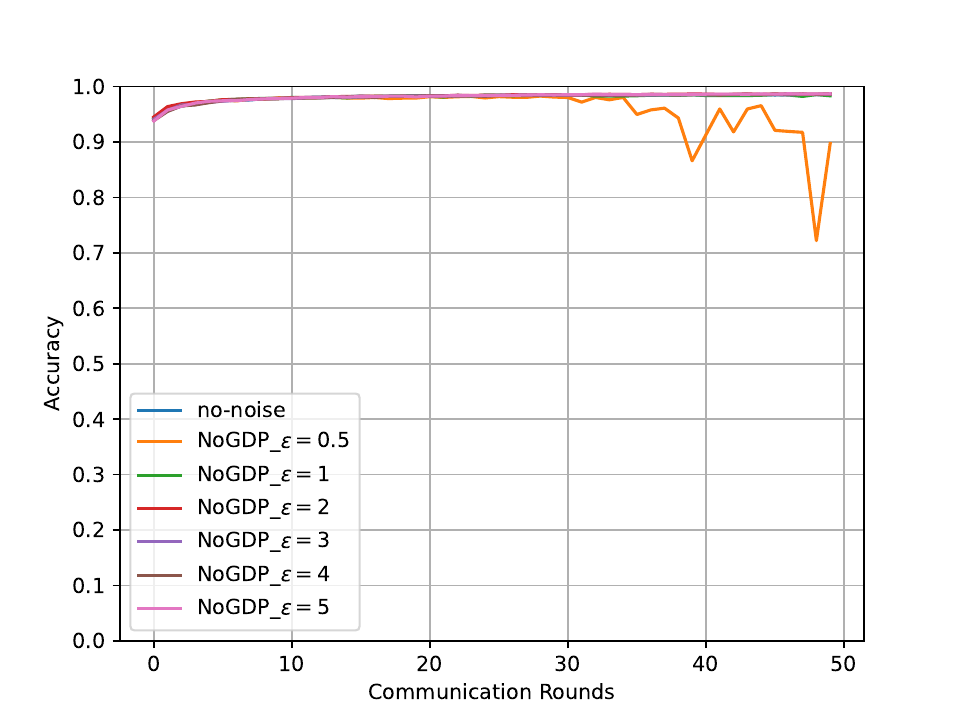}
			\caption{MNIST-IID}
			\label{NoGDPMNIST-IID}
		\end{subfigure} 
		\begin{subfigure}[b]{0.47\textwidth}
			\centering
			\includegraphics[width=\textwidth]{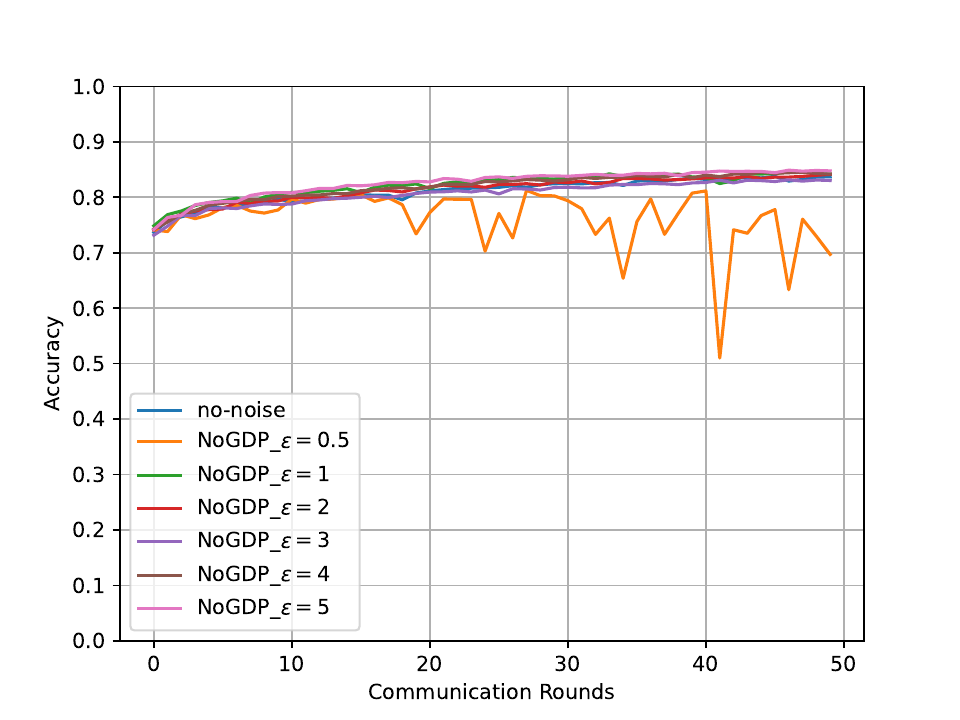}
			\caption{Fashion-MNIST-IID}
			\label{NoGDPFashion-MNIST-IID}
		\end{subfigure}
		\caption{Accuracy of PPBFL under Different $\epsilon$ for MNIST and Fashion-MNIST task, only add local differential privacy noise, $\epsilon = \{0.5, 1, 2, 3, 4, 5\}$}
		\label{Accuracy-NoGDP}
	\end{figure}
	
	
	Building upon CAFL, we have introduced a novel perturbation equation by replacing $\mu$ with the distance between weights and the central point of the corresponding layer in the $\mathbb{GM}$, transitioning from Equation \ref{CAFLM} to Equation \ref{POMEGA}. When solely incorporating differential privacy noise at $\mathbb{LT}$ without introducing $\mathbb{GM}$ differential privacy noise, the model accuracy is depicted in Figure \ref{Accuracy-NoGDP}.
	
	When applying local model differential privacy noise exclusively on the MNIST dataset, as illustrated in the results for $\epsilon$ set at 0.5, the model's accuracy experiences fluctuations after 30 rounds. However, with $\epsilon$ values surpassing 0.5, the model's accuracy aligns with scenarios where no noise is added. Similarly, on the Fashion-MNIST dataset, at $\epsilon$ set to 0.5, the model's accuracy fluctuates and is lower than instances with larger $\epsilon$ values. Yet, when $\epsilon$ exceeds 0.5, the model's accuracy results closely resemble scenarios without added noise.
	
	In comparison to CAFL, our proposed  PPBFL exhibits improved noise tolerance when local model differential privacy noise is exclusively added. Under the same $\epsilon$ conditions, such as $\epsilon$ set at 0.5 or 1, PPBFL outperforms CAFL. 
	The results indicate that our method achieved better model performance while ensuring differential privacy.
	This enhancement stems from the introduced weight distance calculation method, perturbation equation, and the probability distribution of the Bernoulli variable, which, while maintaining the same guarantees of differential privacy security, reduces the addition of differential privacy noise. This reduction contributes to superior model performance under the given conditions.

\subsubsection{Accuracy of PPBFL }
	\begin{figure*}[htbp]
		\centering
	\begin{subfigure}[b]{0.47\textwidth}
		\centering
		\includegraphics[width=\textwidth]{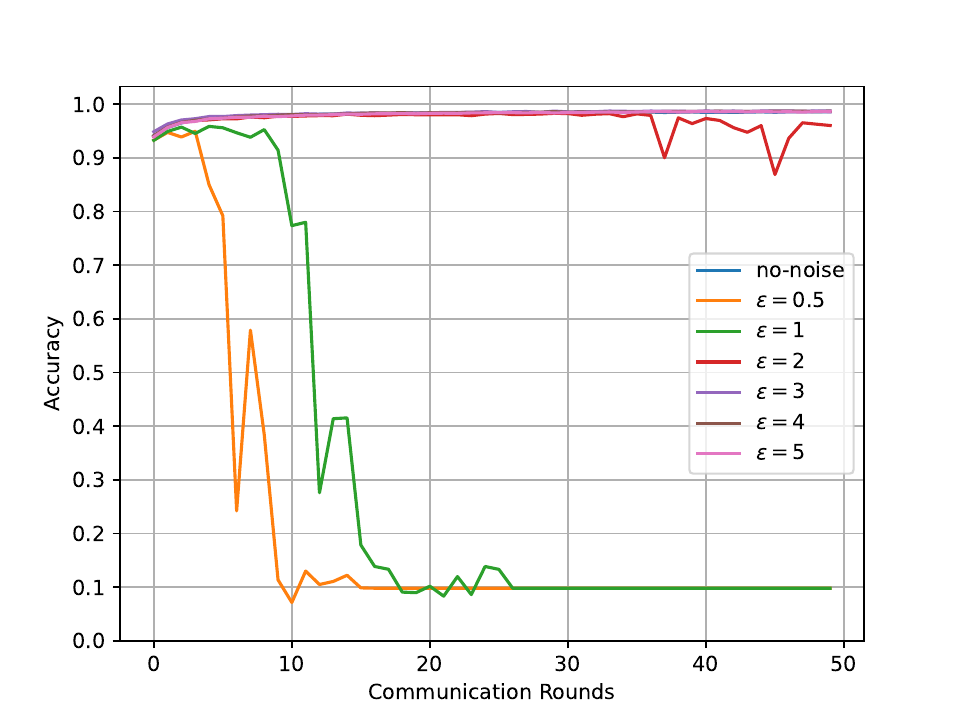}
		\caption{MNIST-IID}
		\label{MNIST-IID}
	\end{subfigure}
	\begin{subfigure}[b]{0.47\textwidth}
	\centering
	\includegraphics[width=\textwidth]{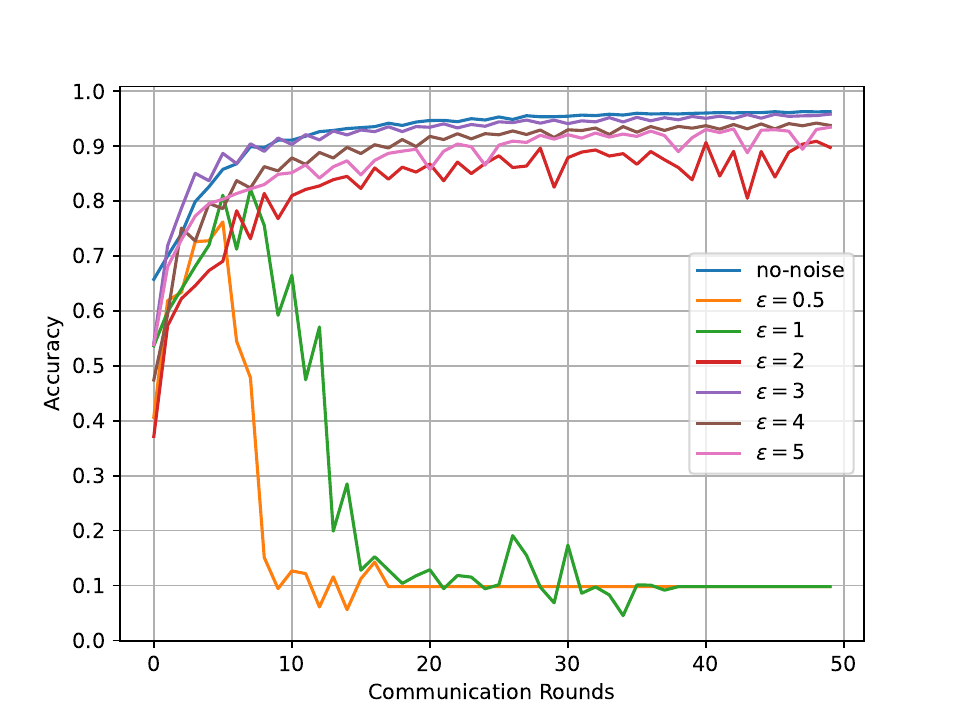}
	\caption{MNIST-Non-IID}
	\label{MNIST-Non-IID}
	\end{subfigure}
	
	\begin{subfigure}[b]{0.47\textwidth}
		\centering
		\includegraphics[width=\textwidth]{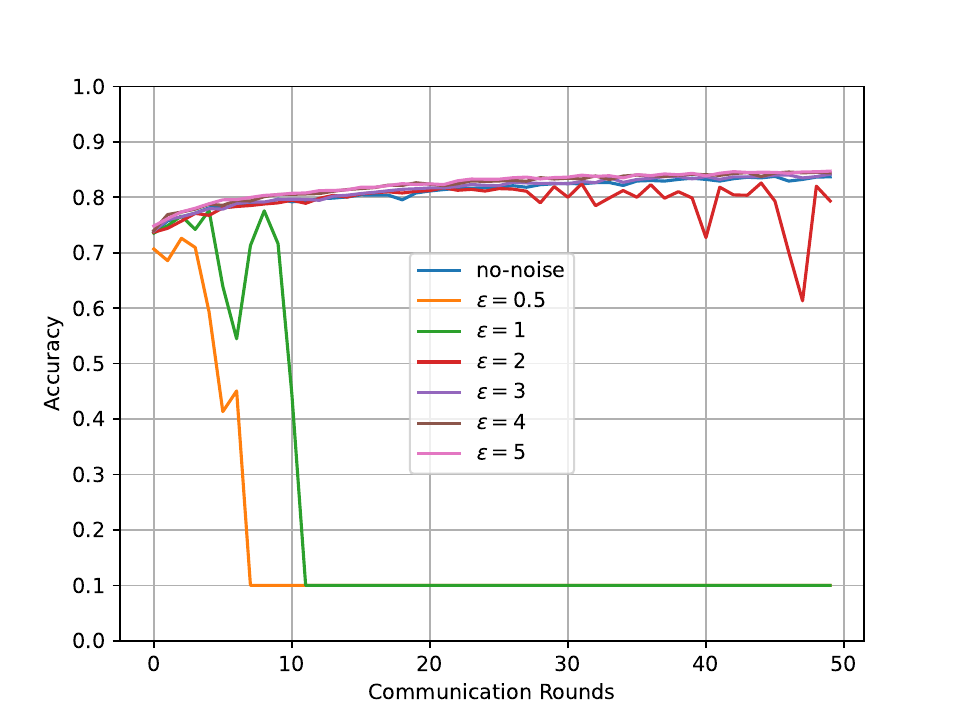}
		\caption{Fashion-MNIST-IID}
		\label{Fashion-MNIST-IID}
	\end{subfigure} 
	\begin{subfigure}[b]{0.47\textwidth}
		\centering
		\includegraphics[width=\textwidth]{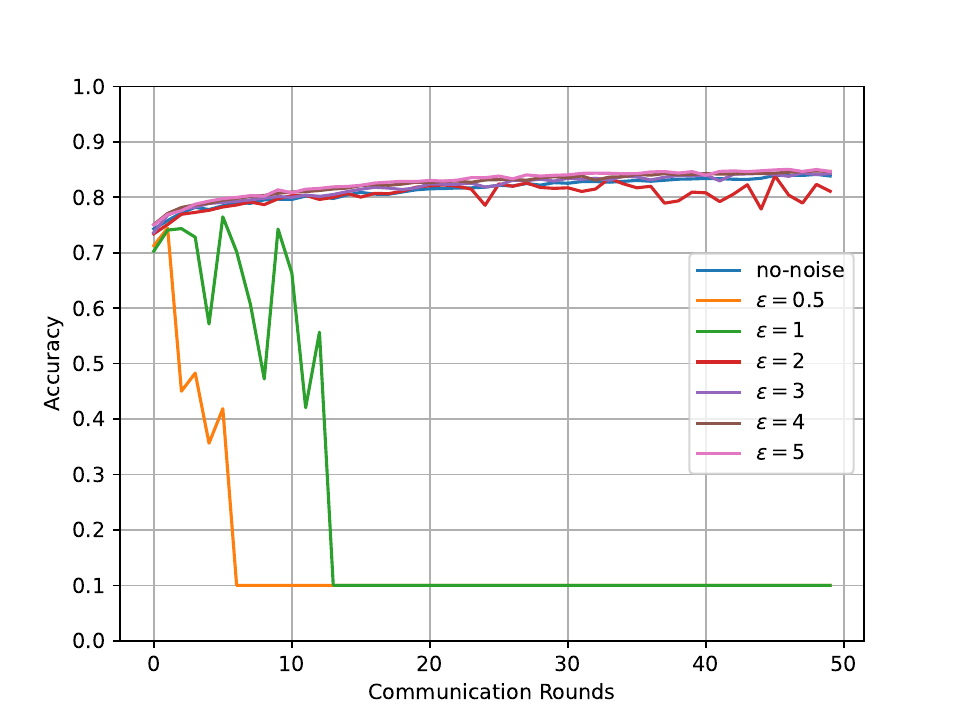}
		\caption{Fashion-MNIST-Non-IID}
		\label{Fashion-MNIST-Non-IID}
	\end{subfigure}
	\caption{Accuracy of PPBFL under Different $\epsilon$ for MNIST and Fashion-MNIST task, $\epsilon = \{0.5, 1, 2, 3, 4, 5\}$}
	\label{Accuracy}
\end{figure*}
%

In Figure \ref{Accuracy}, we conducted tests on the accuracy of PPBFL under different data distributions for MNIST and Fashion-MNIST datasets. By adjusting various values of $\epsilon$, we obtained comparative results for model accuracy. Figure \ref{MNIST-IID} illustrates the training outcomes on the MNIST dataset under the condition of data being independently and identically distributed (IID). It can be observed that when $\epsilon$ is set to 0.5 and 1, the model's accuracy decreases to around 0.1 within the initial 20 rounds, maintaining this accuracy in the subsequent iterations. This phenomenon occurs because, with a smaller $\epsilon$, the privacy protection enforced by differential privacy is more stringent, leading to larger added noise and, consequently, poorer model performance. For $\epsilon$ equal to 2, the model's accuracy aligns with the non-noise-added model for the first 30 rounds, but experiences fluctuations after the 30th round, with accuracy dropping below that of the non-noise-added model at the 50th round.

Comparing this with Figure \ref{NoGDPMNIST-IID}, we observe that the fluctuation in the later rounds of the model with $\epsilon$ set to 2 in Figure \ref{MNIST-IID} is attributed to the addition of differential privacy noise not only to the $\mathbb{LM}$ but also to the $\mathbb{GM}$, impacting the model due to late-stage noise. Moving on to Figure \ref{MNIST-Non-IID}, \ref{Fashion-MNIST-IID}, and \ref{Fashion-MNIST-Non-IID}, similar trends are observed, with the model's accuracy decreasing to around 0.1 within the initial 20 rounds for $\epsilon$ values of 0.5 and 1. However, when $\epsilon$ is set to 2, the model exhibits fluctuations in accuracy during the later training stages.
 
For larger values of $\epsilon$, the model's accuracy remains comparable to the non-noise-added model when training data is independently and identically distributed. However, in the case of non-identically distributed training data, particularly on the MNIST dataset, the addition of differential privacy noise adversely affects the model's performance, causing it to be less accurate compared to the non-noise-added model. This discrepancy arises because in non-identically distributed data, there may be stronger correlations, amplifying the impact of added differential privacy noise on model performance.

In comparison to CAFL, our model PPBFL, although exhibiting fluctuations in later training stages with $\epsilon$ set to 2, generally maintains a higher overall performance. Additionally, we introduce differential privacy noise to the $\mathbb{GM}$, enhancing the security of FL model parameters and mitigating security risks associated with a large number of $\mathbb{LT}$ in the FL process due to the \textit{composition theorem}.

\subsection{Accuracy and Stakes in PPBFL and PoS Consensus algorithm}

\begin{figure}[htbp]
	\centering
	\begin{subfigure}[b]{0.47\textwidth}
		\centering
		\includegraphics[width=\textwidth]{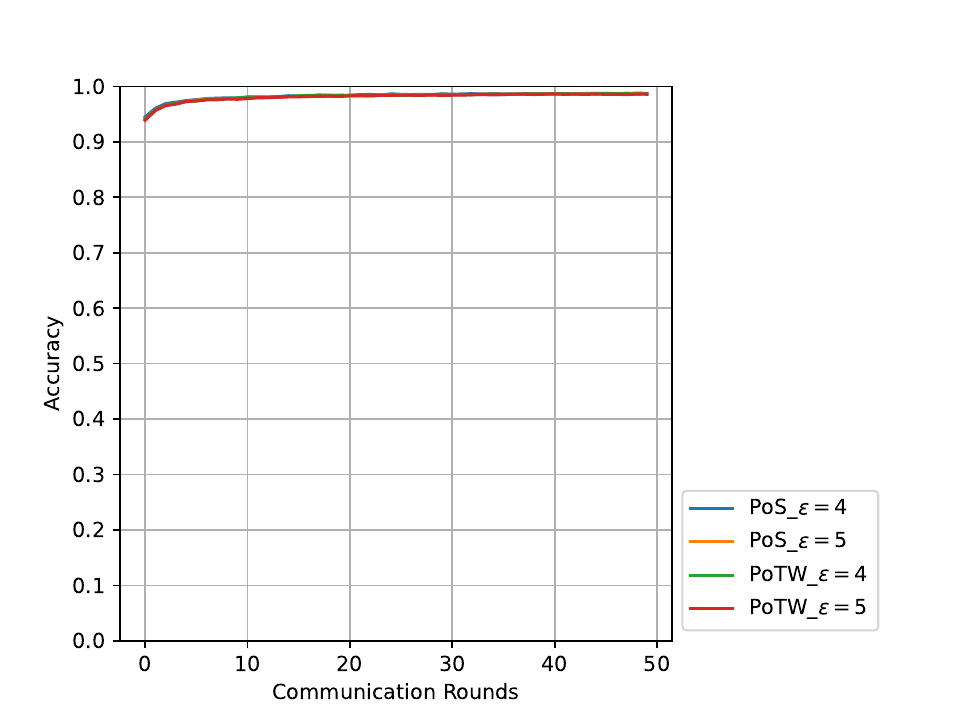}
		\caption{MNIST-IID}
		\label{Consensus Algorithms MNIST-IID}
	\end{subfigure}
	\begin{subfigure}[b]{0.47\textwidth}
	\centering
	\includegraphics[width=\textwidth]{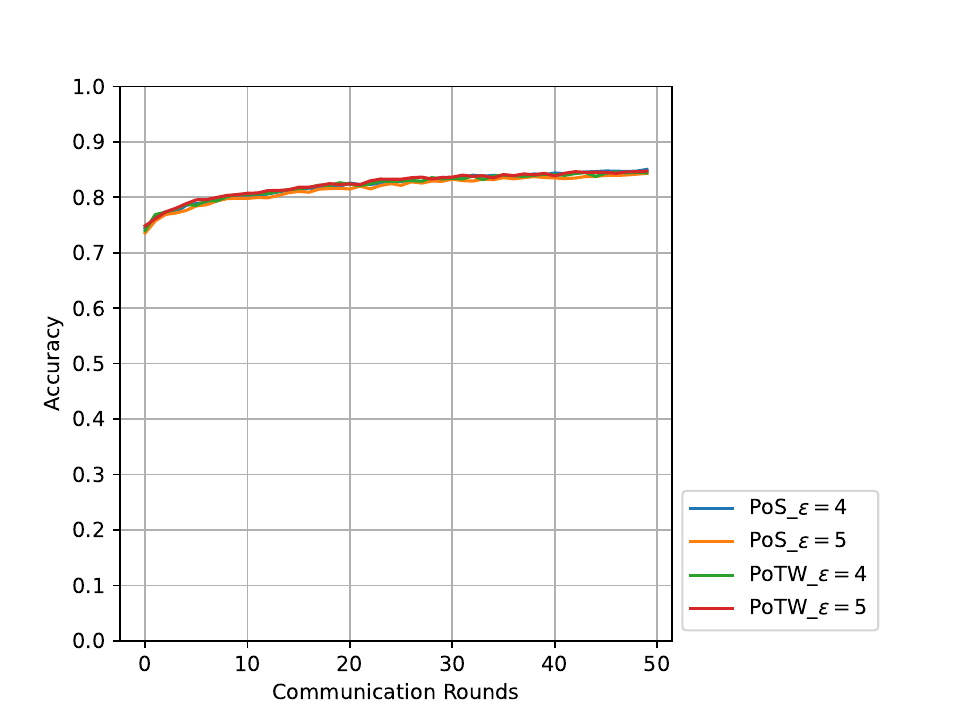}
	\caption{Fashion-MNIST-IID}
	\label{Consensus Algorithms Fashion-MNIST-IID}
	\end{subfigure} 
	\caption{Accuracy of PPBFL under Different $\epsilon$ and  Consensus Algorithms for MNIST and Fashion-MNIST task, $\epsilon = \{4, 5\}$ }
	\label{Consensus Algorithms Accuracy}
\end{figure}

\begin{figure}[htbp]
	\centering

	\begin{subfigure}[b]{0.24\textwidth}
		\centering
		\includegraphics[width=\textwidth]{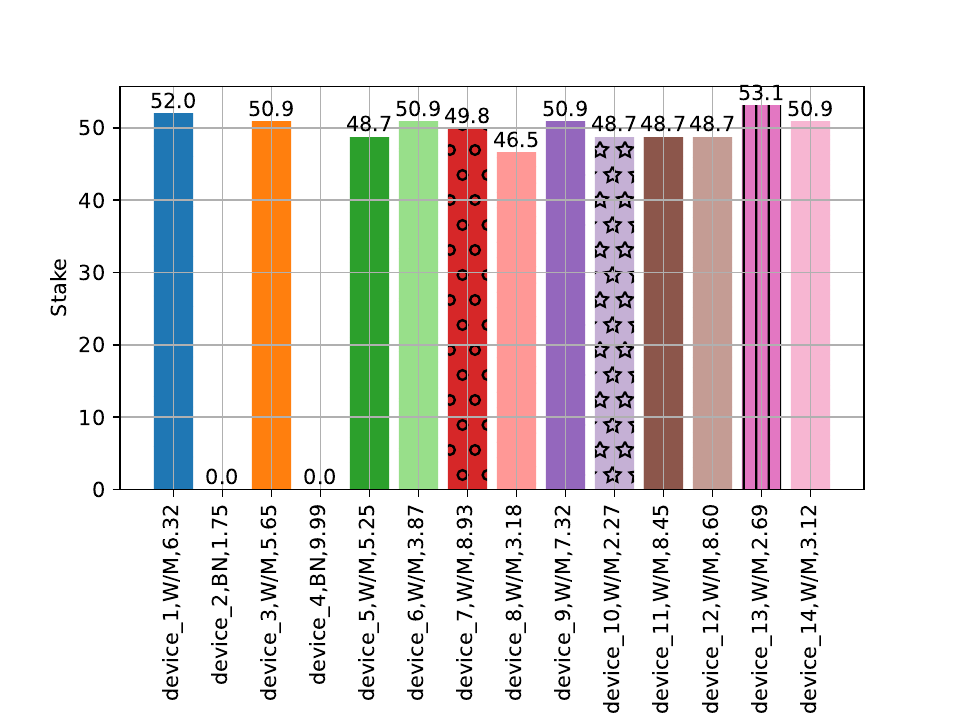}
		\caption{MNIST-PoS, $\epsilon = 4$}
		\label{MNIST-PoS4}
	\end{subfigure} 
	\begin{subfigure}[b]{0.24\textwidth}
		\centering
		\includegraphics[width=\textwidth]{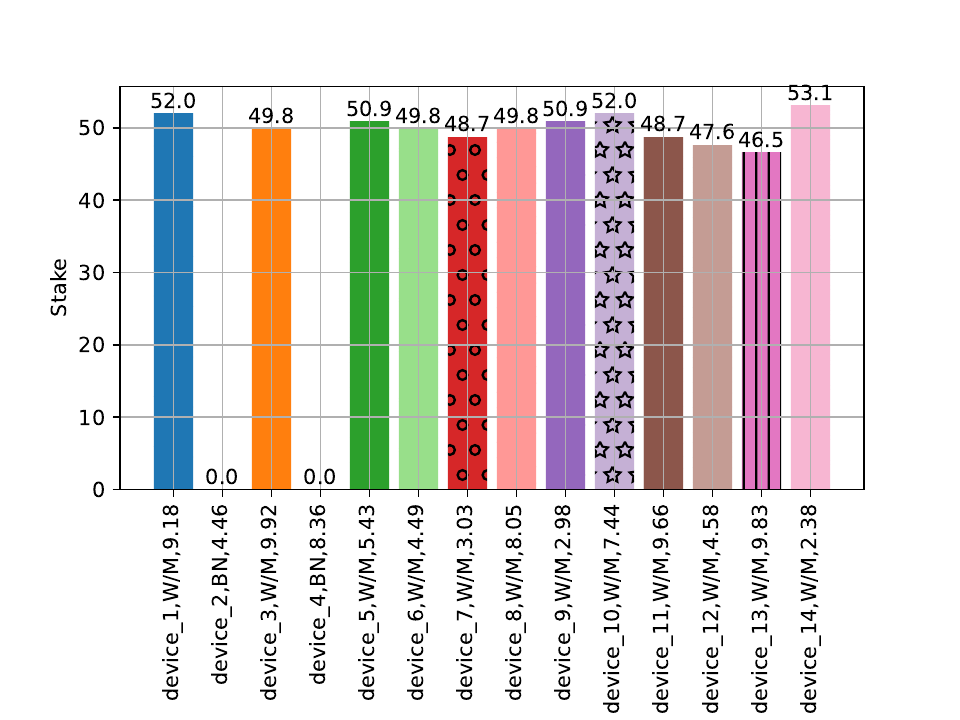}
		\caption{MNIST-PoS, $\epsilon = 5$}
		\label{MNIST-PoS5}
	\end{subfigure}

	\begin{subfigure}[b]{0.24\textwidth}
		\centering
		\includegraphics[width=\textwidth]{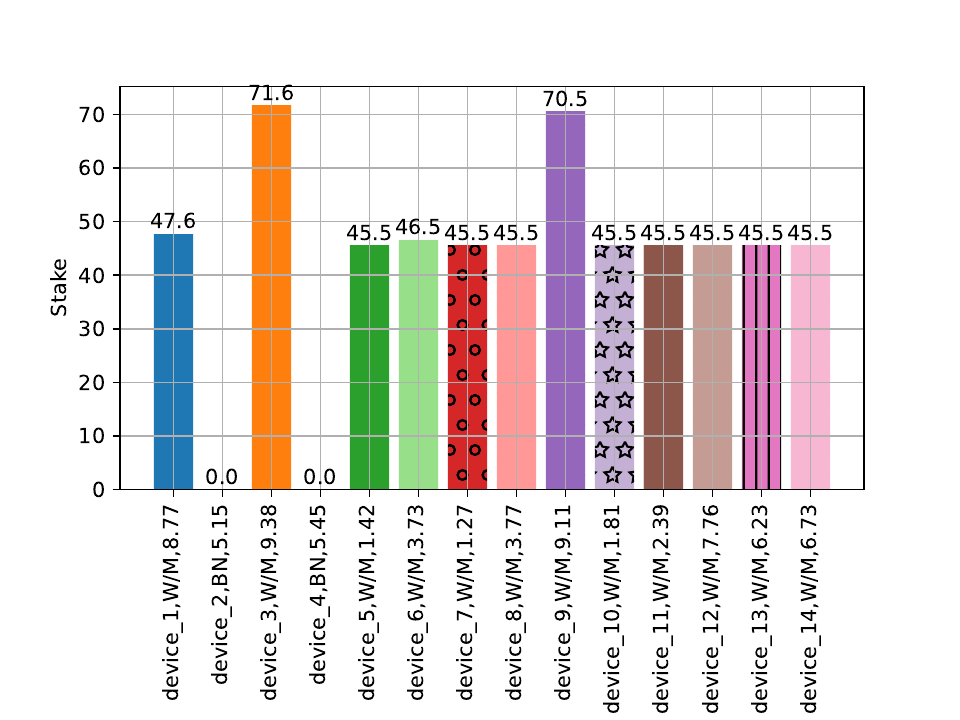}
		\caption{MNIST-PoTW, $\epsilon = 4$}
		\label{MNIST-PoTW4}
	\end{subfigure}
	\begin{subfigure}[b]{0.24\textwidth}
		\centering
		\includegraphics[width=\textwidth]{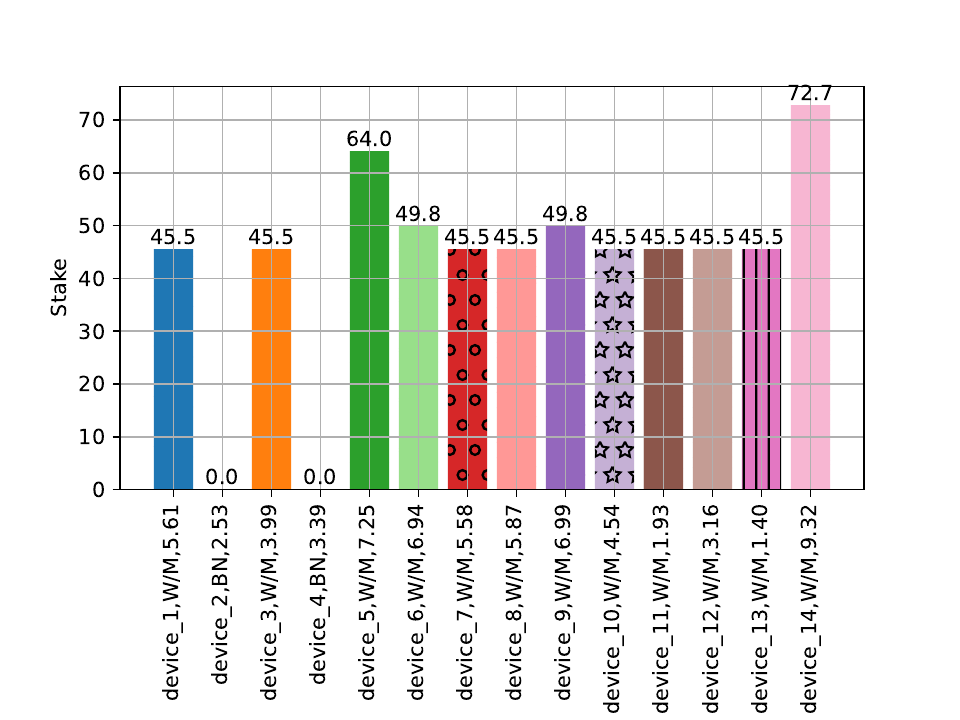}
		\caption{MNIST-PoTW, $\epsilon = 5$}
		\label{MNIST-PoTW5}
	\end{subfigure}
	
	\begin{subfigure}[b]{0.24\textwidth}
		\centering
		\includegraphics[width=\textwidth]{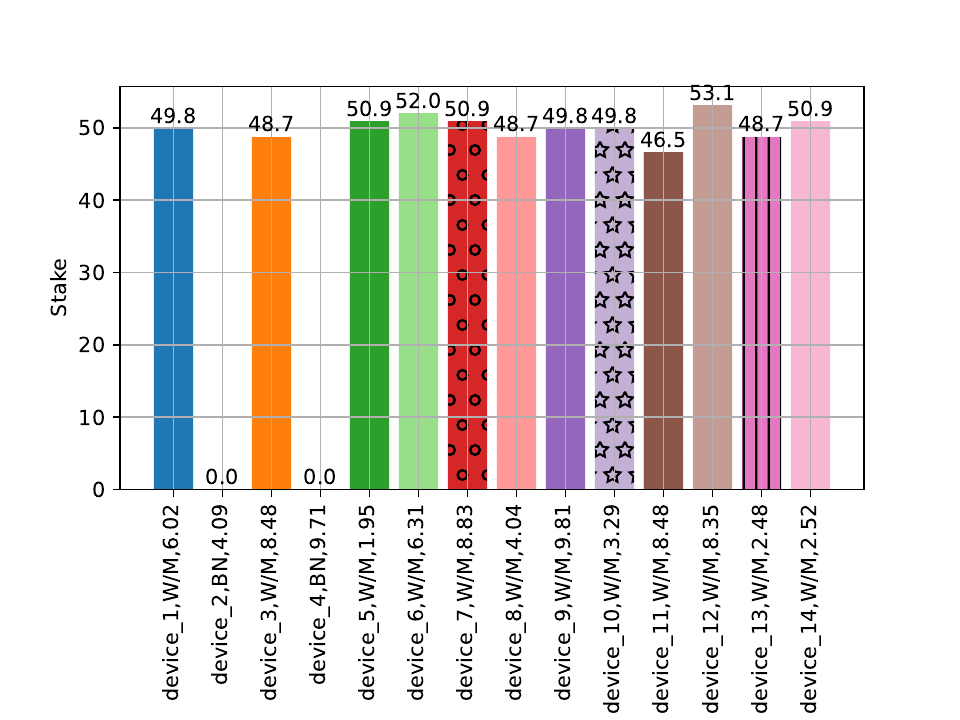}
		\caption{Fashion-MNIST-PoS, $\epsilon = 4$}
		\label{Fashion-MNIST-PoS4}
	\end{subfigure} 
	\begin{subfigure}[b]{0.24\textwidth}
		\centering
		\includegraphics[width=\textwidth]{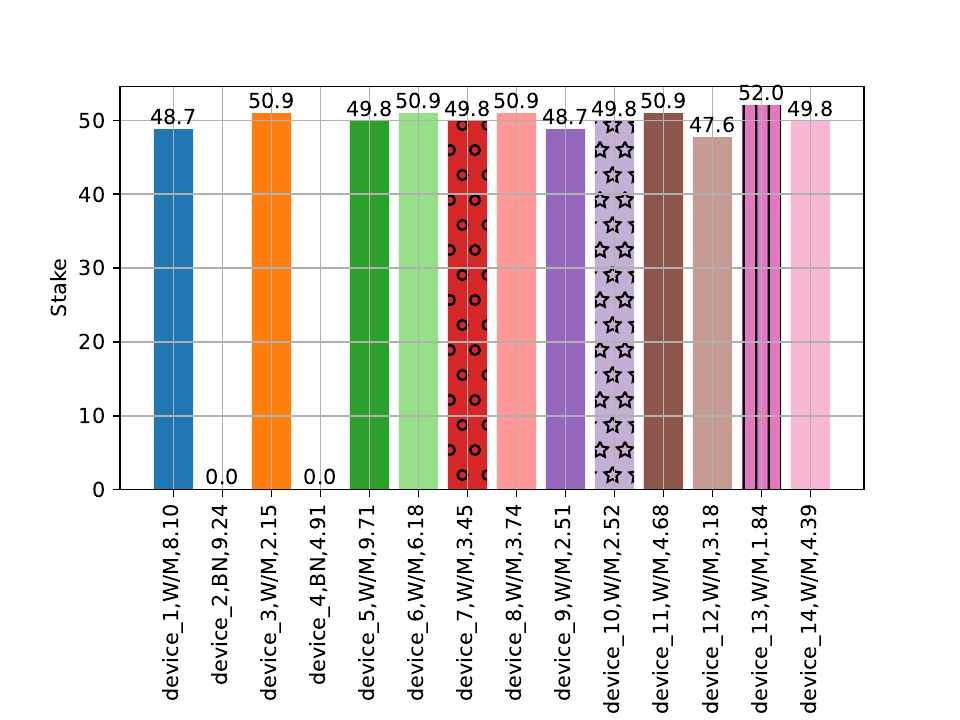}
		\caption{Fashion-MNIST-PoS, $\epsilon = 5$}
		\label{Fashion-MNIST-PoS5}
	\end{subfigure}

	\begin{subfigure}[b]{0.24\textwidth}
		\centering
		\includegraphics[width=\textwidth]{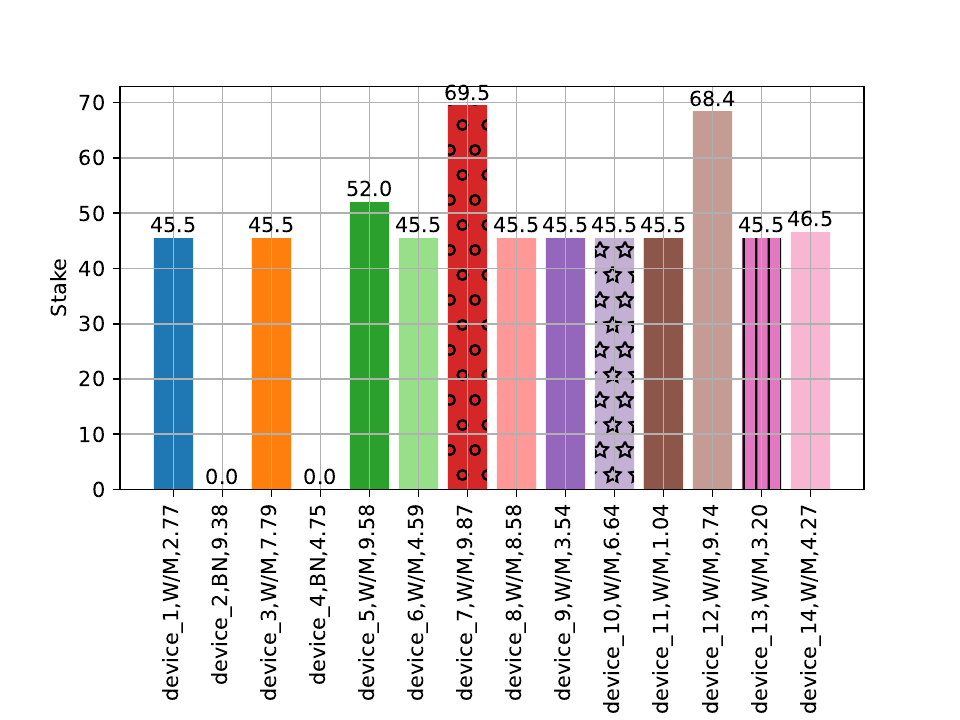}
		\caption{Fashion-MNIST-PoTW, $\epsilon = 4$}
		\label{Fashion-MNIST-PoTW4}
	\end{subfigure}
	\begin{subfigure}[b]{0.24\textwidth}
		\centering
		\includegraphics[width=\textwidth]{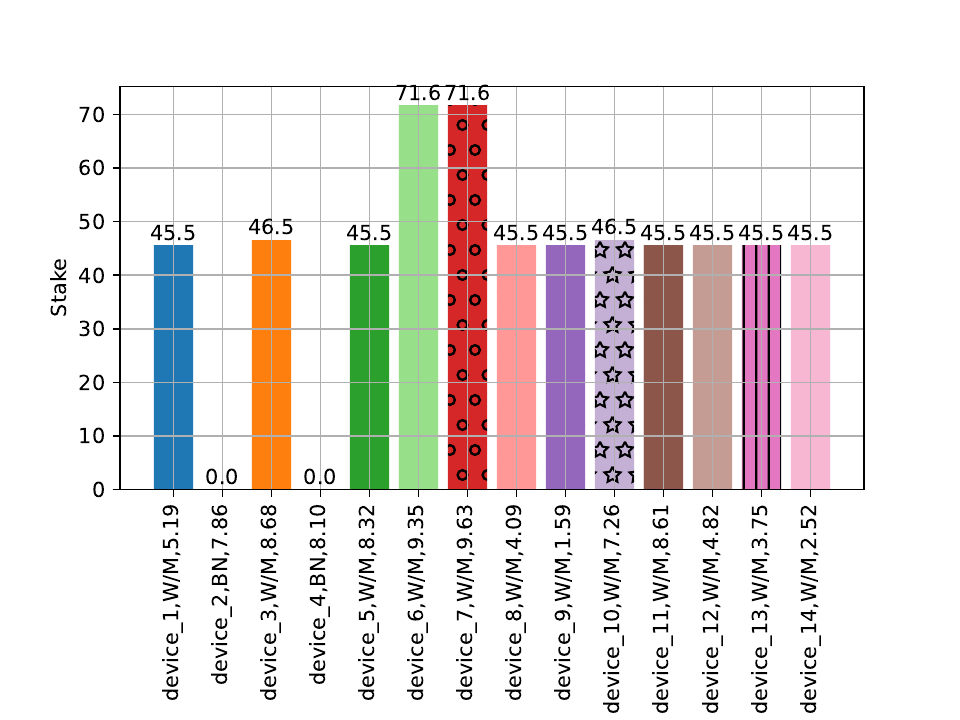}
		\caption{Fashion-MNIST-PoTW, $\epsilon = 5$}
		\label{Fashion-MNIST-PoTW5}
	\end{subfigure}
		
	\caption{Stake of Different Devices for PoS and PoTW Consensus Algorithms under MNIST and Fashion-MNIST task, $\epsilon = \{4, 5\}$}, 
	\label{Stake}
\end{figure}

%

Figures \ref{Consensus Algorithms Accuracy} and \ref{Stake} illustrate a comparison between the PoS consensus algorithm and our proposed PoTW consensus algorithm in terms of model performance and node stake on the MNIST and Fashion-MNIST datasets. The PoS consensus algorithm selects the node with the maximum coin age as the packing node, while our proposed PoTW consensus algorithm utilizes the completion time of $\mathbb{LT}$ for FL tasks as proof of workload during the FL process, selecting the node with the fastest completion time as the packing node. We simulate different node capabilities, where in Figure \ref{Stake}, the x-axis labels, starting with 'Device\_' and followed by numbers, denote different devices. 'W/M' signifies that the device participates in FL training, possibly as a local training node or a packing node. "BN" indicates that the device exists in the blockchain but does not participate in the FL training process, and hence, the consensus algorithm does not distribute stake to it.

Figure \ref{Consensus Algorithms MNIST-IID} displays the performance of PPBFL on the MNIST dataset using independently and identically distributed training data, employing PoS and PoTW consensus algorithms, and $\epsilon = \{4, 5\}$. Figure \ref{Consensus Algorithms Fashion-MNIST-IID} showcases the performance of PPBFL on the Fashion-MNIST dataset with independently and identically distributed training data. From Figure \ref{Consensus Algorithms Accuracy}, it is observed that different consensus algorithms exhibit consistent model performance on the same dataset. The choice of different consensus algorithms does not significantly impact the model performance for FL training tasks. This is because consensus algorithms do not participate in the training process of FL; they merely reward $\mathbb{LT}$ with strong computational capabilities to facilitate the normal progression of FL tasks, and therefore do not influence the quality of FL model performance.

Figure \ref{Stake} illustrates the stake obtained by different nodes at the 50th round when using PoS and PoTW consensus algorithms on the MNIST and Fashion-MNIST datasets. From Figure \ref{Stake}, it can be observed that when using the PoS consensus algorithm, the stake obtained by different nodes is similar. In contrast, when using the PoTW consensus algorithm, nodes with higher computational capabilities receive more stake. This is because the PoTW consensus algorithm selects the node with the strongest computational power in the current round as the packing node, which earns more rewards. At the same time, the packing node does not participate in the current round's model training, making it ineligible to be selected as the packing node in the next round. In the PoS consensus algorithm, the packing node is chosen based on the node's coin age. When the coin ages of different nodes are similar, the likelihood of different nodes becoming the packing node is also similar. Therefore, in the PoS consensus algorithm, the stake amounts of different nodes are not identical but are similar. In the PoTW consensus algorithm, the nodes with the highest computational power in the first few nodes receive the most stake, while the stake of the remaining nodes remains consistent.

\section{Conclusion}
\label{Conclusion}

In this article, we propose PPBFL to protect model parameter privacy and enhance the participation of clients in model training in FL. We introduce a dual adaptive differential privacy addition mechanism, which involves adding adaptive differential privacy noise to both $\mathbb{LM}$ and $\mathbb{GM}$. During the process of adding differential privacy noise, we introduce the zero-bias noise proposed by us. This not only prevents inference attacks but also addresses the security degradation issue that arises when multiple $\mathbb{LM}$ with added differential privacy noise are combined into the $\mathbb{GM}$. By combining ring signatures and introducing the mix transactions mechanism, we safeguard the identity privacy of $\mathbb{LT}$. Additionally, we present the proof of training work, addressing the resource wastage problem in the Proof of Work (PoW) consensus algorithm by transforming the computational challenge into model training, thereby increasing node motivation. Security analysis demonstrates that our proposed PPBFL exhibits high security, and experiments show that our proposed approach yields favorable model performance.

\section*{Acknowledgments}
This work was supported by the National Natural Science Foundation of China under Grant No. 62272024.




\section{Biography Section}
 
\vspace{11pt}

\vspace{-33pt}
\begin{IEEEbiography}[{\includegraphics[width=1in,height=1.25in,clip,keepaspectratio]{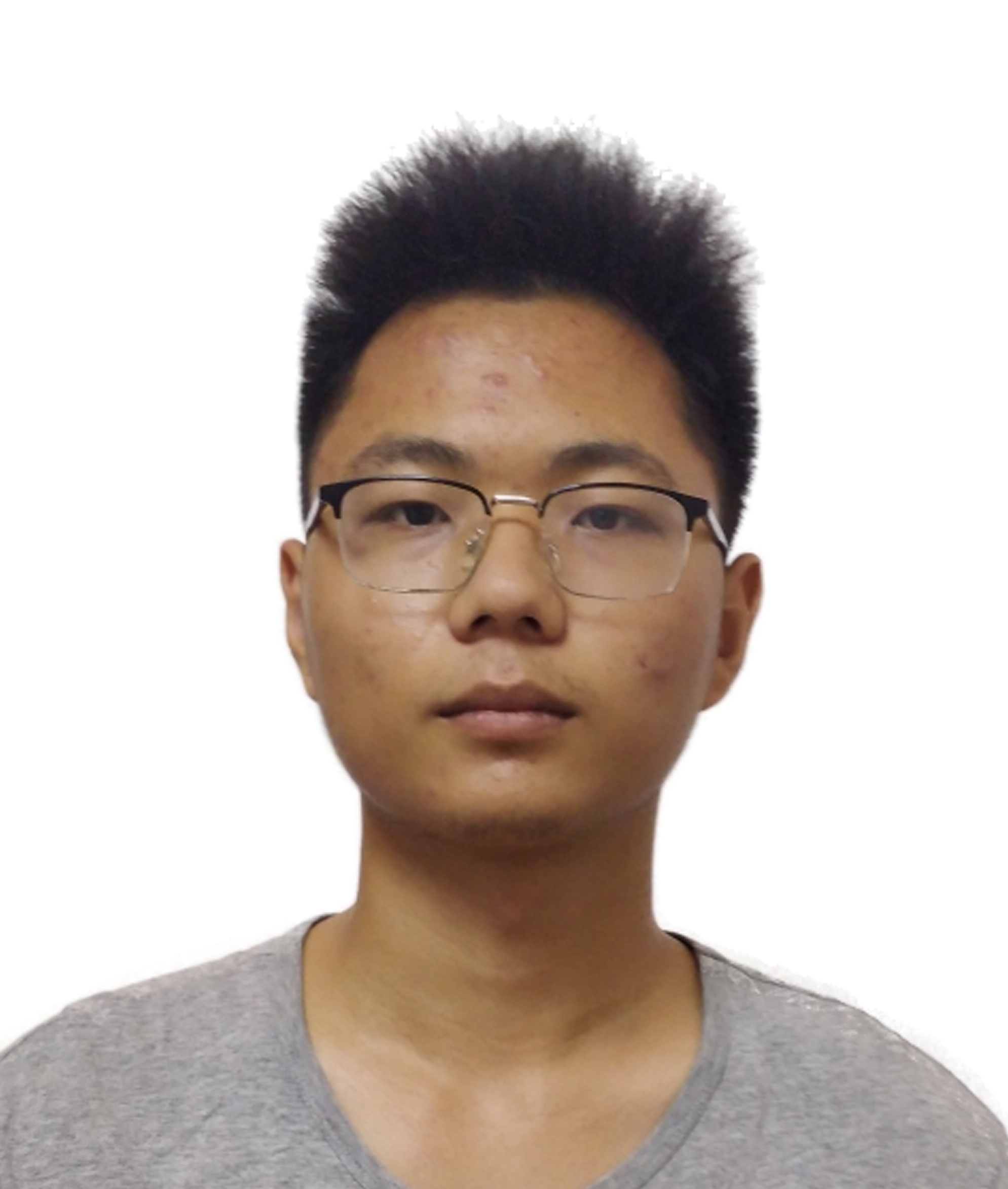}}]{Yang Li}
 received the master’s degree in computer technology from  Zhengzhou University, Zhengzhou, China, in 2021. He is
currently pursuing the Ph.D. degree with Beihang
University, Beijing, China, under the supervision of
Prof. Chunhe Xia.
His research interests include  federated learning and blockchain.
\end{IEEEbiography}

\vspace{11pt}

\begin{IEEEbiography}[{\includegraphics[width=1in,height=1.25in,clip,keepaspectratio]{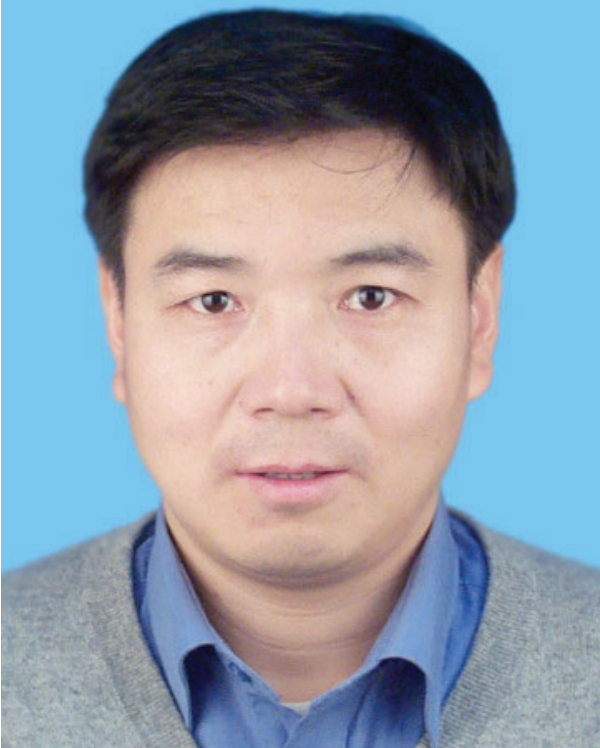}}]{Chunhe Xia}
	received the Ph.D. degree in computer application from Beihang University, Beijing, China, in 2003.
	
	He is currently a Supervisor and a Professor with Beihang University, where he is also the Director of the Beijing Key Laboratory of Network Technology and a Professor of the Guangxi Collaborative Innovation Center of Multi-Source Information Integration and Intelligent Processing. He has participated in different national major research projects and has published more than 70 research papers in important international conferences and journals. His current research focuses on network and information security, information 	countermeasure, cloud security, and network measurement.
\end{IEEEbiography}

\vspace{11pt}
\begin{IEEEbiography}[{\includegraphics[width=1in,height=1.25in,clip,keepaspectratio]{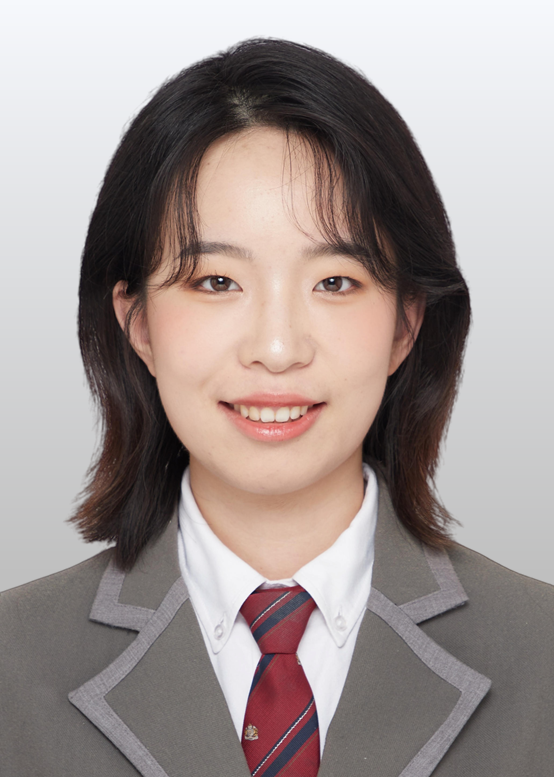}}]{Wanshuang Lin}
	received the B.E. degree in Information Security from the China University of Geosciences (Wuhan), Wuhan, China, in 2019, and is currently pursuing the Ph.D. degree with the School of Cyber Science and Technology, Beihang University, Beijing, China. Her primary research interests lie in network security and machine learning.
\end{IEEEbiography}
\vspace{11pt}
\begin{IEEEbiography}[{\includegraphics[width=1in,height=1.25in,clip,keepaspectratio]{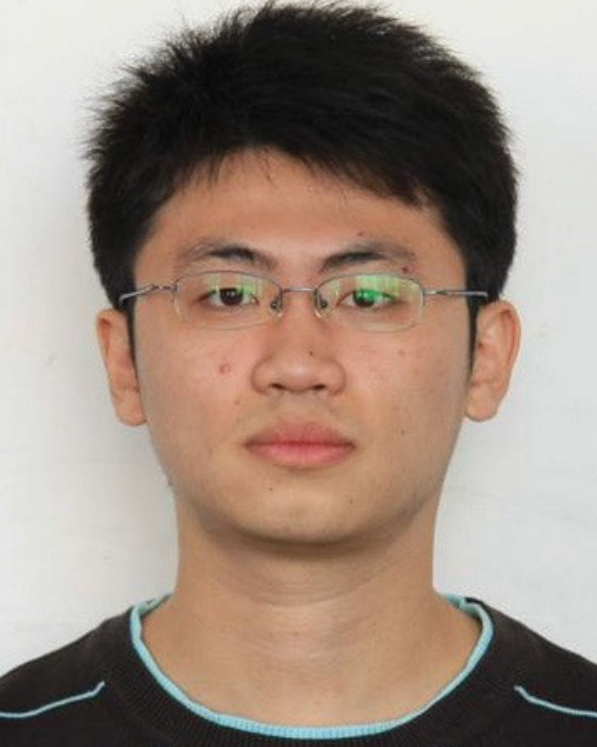}}]{Tianbo Wang}
(Member, IEEE) received the Ph.D. degree in computer application from Beihang University, Beijing, China, in 2018.

He is currently an Associate Professor with Beihang University, where he is also an Associated Professor of the Shanghai Key Laboratory of Computer Software Evaluating and Testing. He has participated in several National Natural Science Foundations and other research projects. His research interests include network and information security, intrusion detection technology, and information countermeasure.
\end{IEEEbiography}

%

\vfill

\end{document}